\documentclass[11pt,a4paper]{article}
\usepackage{amsmath,amsfonts,amssymb,amsthm,amstext,amscd,array,bbold,eucal}
\usepackage[all]{xy}

\DeclareMathOperator{\Sphere}{S}
\DeclareMathOperator{\Torus}{T}

\DeclareMathOperator{\Conn}{Conn}
\let\S\Sphere

\newcommand{\ee}[1]{{\rm e}^{#1}}
\newcommand{\ii}{{\rm i}}

\def\Dirac{{D\!\!\!\!/\,}} 

\newcommand{\mbf}[1]{{\boldsymbol {#1} }}
\def\ii{{\,{\rm i}\,}}
\def\dd{{\rm d}}

\def\YM{{\rm YM}}
\def\Ext{{\rm Ext}}
\def\P{{\sf P}}

\def\K{{\rm K}}
\def\T{{\sf T}}

\def\shift{{\sf S}}

\def\mfg{{\mathfrak g}}

\def\mcS{{\mathcal S}}

\def\compact{{\mathcal K}}

\def\Fock{{\mathcal F}}
\def\Hil{{\mathcal H}}
\def\alg{{\mathcal A}}
\def\lin{{\mathcal B}}
\def\Calkin{{\mathcal Q}}

\newcommand{\eq}{\begin{equation}}
\newcommand{\eqend}{\end{equation}}
\newcommand{\eqa}{\begin{eqnarray}}
\newcommand{\nonueqa}{\begin{eqnarray*}}
\newcommand{\eqaend}{\end{eqnarray}}
\newcommand{\nonueqaend}{\end{eqnarray*}}

\newcommand{\bma}[1]{\begin{array}{#1}}
\newcommand{\ema}{\end{array}}
\newcommand{\bc}{\begin{center}}
\newcommand{\ec}{\end{center}}

\renewcommand{\thefootnote}{\fnsymbol{footnote}}
\newcommand{\newsection}{\setcounter{equation}{0}\section}

\def\appendix#1{\addtocounter{section}{1}\setcounter{equation}{0}
\renewcommand{\thesection}{\Alph{section}}
\section*{Appendix \thesection\protect\indent \parbox[t]{11.715cm} {#1}}
\addcontentsline{toc}{section}{Appendix \thesection\ \ \ #1} }

\newcommand{\complex}{{\mathbb C}} 
\newcommand{\zed}{{\mathbb Z}} 
\newcommand{\nat}{{\mathbb N}} 
\newcommand{\real}{{\mathbb R}} 

\newcommand{\rat}{{\mathbb Q}} 
\newcommand{\mat}{{\mathbb M}} 
\newcommand{\plane}{{\mathbb V}}
\newcommand{\id}{{1\!\!1}} 
\def\Dirac{{D\!\!\!\!/\,}} 
\def\alg{{\mathcal A}}
\def\hil{{\mathcal H}}
\def\mod{{\mathcal E}}
\def\Weyl{{\mathcal W}}

\newif\ifold             \oldtrue

\newcommand{\Tr}[1]{\:{\rm Tr}\,#1}

\def\e{{\,\rm e}\,}

\def\be{\begin{equation}}
\def\ee{\end{equation}}
\def\bea{\begin{eqnarray}}
\def\eea{\end{eqnarray}}
\def\bd{\begin{displaymath}}
\def\ed{\end{displaymath}}

\def\one{\mbox{1 \kern-.59em {\rm l}}}

\newcommand{\beq}{\begin{eqnarray}}
\newcommand{\eeq}{\end{eqnarray}}

\makeatletter
\newdimen\normalarrayskip              
\newdimen\minarrayskip                 
\normalarrayskip\baselineskip
\minarrayskip\jot
\newif\ifold             \oldtrue            
\def\arraymode{\ifold\relax\else\displaystyle\fi} 
\def\@arrayskip{\ifold\baselineskip\z@\lineskip\z@
     \else
     \baselineskip\minarrayskip\lineskip2\minarrayskip\fi}
\def\@arrayclassz{\ifcase \@lastchclass \@acolampacol \or
\@ampacol \or \or \or \@addamp \or
   \@acolampacol \or \@firstampfalse \@acol \fi
\edef\@preamble{\@preamble
  \ifcase \@chnum
     \hfil$\relax\arraymode\@sharp$\hfil
     \or $\relax\arraymode\@sharp$\hfil
     \or \hfil$\relax\arraymode\@sharp$\fi}}
\def\@array[#1]#2{\setbox\@arstrutbox=\hbox{\vrule
     height\arraystretch \ht\strutbox
     depth\arraystretch \dp\strutbox
     width\z@}\@mkpream{#2}\edef\@preamble{\halign \noexpand\@halignto
\bgroup \tabskip\z@ \@arstrut \@preamble \tabskip\z@ \cr}%
\let\@startpbox\@@startpbox \let\@endpbox\@@endpbox
  \if #1t\vtop \else \if#1b\vbox \else \vcenter \fi\fi
  \bgroup \let\par\relax
  \let\@sharp##\let\protect\relax
  \@arrayskip\@preamble}
\makeatother

\newtheorem{theorem}{Theorem}[section]
\newtheorem{lemma}{Lemma}[section]
\newtheorem{cor}{Corollary}[section]
\newtheorem{proposition}{Proposition}[section]
\theoremstyle{definition}
\newtheorem{definition}{Definition}[section]
\theoremstyle{remark}
\newtheorem{example}{Example}[section]
\newtheorem{remark}{Remark}[section]

\begin{document}
\begin{flushright}

\baselineskip=12pt

HWM--05--27\\
EMPG--05--17\\
hep--th/yymmnnn\\
\hfill{ }\\
December 2005
\end{flushright}

\begin{center}

\vspace{5mm}

\baselineskip=24pt

{\Large\bf D-BRANES IN NONCOMMUTATIVE FIELD THEORY}\footnote{Based on
  minicourse given at the {\it International Workshop on
    Noncommutative Geometry NCG2005}, Institute for Studies in
  Theoretical Physics and Mathematics (IPM), Tehran, Iran, September 11--22
  2005. To be published in the minicourse volume by World Scientific
  and in IPM Lecture Notes Series.}

\baselineskip=14pt

\vspace{5mm}

{\bf Richard J. Szabo}
\\[3mm]
{\it Department of Mathematics}\\
and\\ {\it Maxwell Institute for Mathematical Sciences\\
Heriot-Watt University\\
Colin Maclaurin Building, Riccarton, Edinburgh EH14 4AS, U.K.}
\\{\tt R.J.Szabo@ma.hw.ac.uk}
\\[10mm]

\end{center}

\begin{abstract}

\baselineskip=12pt

A mathematical introduction to the
classical solutions of noncommutative field theory is presented, with
emphasis on how they may be understood as states of D-branes in Type~II
superstring theory. Both scalar field theory and gauge theory on Moyal
spaces are extensively studied. Instantons in Yang-Mills
theory on the two-dimensional noncommutative torus and the fuzzy sphere
are also constructed. In some instances the connection to D-brane physics
is provided by a mapping of noncommutative solitons into K-homology.

\end{abstract}



{\baselineskip=10pt

\tableofcontents}


\renewcommand{\thefootnote}{\arabic{footnote}} \setcounter{footnote}{0}

\newsection{Introduction and background from string
  theory\label{Intro}}

These lecture notes provide an introduction to some selected topics in
noncommutative field theory that are motivated by the interaction
between string theory and noncommutative geometry. The material is
intended to be geared at an audience consisting of graduate students
and beginning postdoctoral researchers in mathematics, and in general
any mathematician interested in how string theory crops up in certain
mathematical settings. The material also serves as an introduction to
physicists into some of the more formal aspects of noncommutative
field theory, revolving primarily around the geometric structure of their
classical solutions and the mathematical interpretation of
noncommutative solitons as D-branes.

The main theme that we will focus on revolves around
the scenario that string theory with D-branes in the presence of
``background fields'' leads to noncommutative geometries on the
worldvolumes of the branes. Some familiarity with this notion along
with the basics of noncommutative geometry, the standard examples
of noncommutative spaces, and the rudiments of K-theory are assumed,
as they were discussed in other
minicourses at the workshop. Nevertheless, to set the stage and
terminology, we will take a short mathematical tour in this section
through this story and define the relevant concepts in the
correspondence to make our presentation
as self-contained as possible. Throughout we will deal only with {\it
  classical} aspects of string theory and noncommutative field theory,
and all of our definitions and explanations are to be understood in
this context alone. Quantization introduces many complicated technical
aspects that are out of the scope of these introductory notes.

We begin by explaining some basic concepts from classical string
theory. Let $X$ be an oriented Euclidean
spin manifold with Riemannian metric $G$. We call $X$ the {\it
  spacetime} or the {\it target space}. Let $\Sigma$ be a
Riemann surface. A {\it string} is a harmonic map $x:\Sigma\to X$,
i.e. an immersion of the surface $\Sigma$ in $X$ of minimal area in
the induced metric $x^*(G)$. The surface $\Sigma$ is called the
{\it worldsheet} of the string. The harmonic property can be described in
the usual way by a variational principle based on a $\sigma$-model on
$\Sigma$ with target space $X$. The string is said to be of {\it
  Type~II} if $\Sigma$ is oriented, and of {\it Type~I} if $\Sigma$ is
non-orientable. In the following we will only deal with Type~II
strings. A string is {\it closed} if its worldsheet is closed,
$\partial\Sigma=\emptyset$, while it is {\it open} if its worldsheet
has boundary, $\partial\Sigma\neq\emptyset$. The spin condition on $X$
is assumed so that we can define spinors and ultimately
{\it superstrings}, but we shall not go into any aspects of
supersymmetry here.

In the simplest instance, a {\it D-brane} may be defined as a closed
oriented submanifold $W\subset X$ which can be used as a boundary
condition for open strings. This means that in the presence of
D-branes the admissible open strings are the {\it relative} harmonic
maps $x:(\Sigma,\partial\Sigma)\to(X,W)$. The submanifold $W$ is
called the {\it worldvolume} of the D-brane. Not all submanifolds are
allowed as viable D-brane worldvolumes. For instance, a consistent
choice of boundary conditions must preserve the fundamental conformal
invariance of the string theory. Determining the allowed D-branes in a
given spacetime $X$ is an extremely difficult problem which requires
having the quantum field theory of the worldsheet $\sigma$-model under
control. For example, the cancellation of worldsheet anomalies for
Type~II strings requires $W$ to be a spin$^c$
manifold~\cite{FreedWit1}.

Let us next introduce the concept of a {\it supergravity
  background field}. In addition to the metric $G$ on $X$, we assume
the presence of an additional geometrical entity called the {\it
  Neveu-Schwarz $B$-field}. It is a two-form $B\in\bigwedge^2(T^*X)$,
which we will locally regard as a skew-symmetric linear map
$B_p:T_pX\to T_p^*X$ for $p\in X$. The $B$-field has curvature $H=\dd
B\in\bigwedge^3(T^*X)$ and characteristic class $[H]\in{\rm
  H}^3(X,\zed)$. Of course as $H=\dd B$ is an exact three-form it
defines a trivial class in de~Rham cohomology, but
there can be torsion and other effects which yield a non-trivial
characteristic class. The $H$-field is constrained to obey the {\it
  supergravity equation}
\beq
R(G)=\mbox{$\frac14$}\,H\circ G^{-1}\,\neg\,H \ ,
\label{SUGRAeqs}\eeq
where $R(G)$ is the Ricci curvature two-form of the metric $G$ and
$\neg$ denotes contraction. Similarly to the $B$-field, the metric
here and throughout is regarded locally as a symmetric non-degenerate
homomorphism $G_p:T_pX\to T_p^*X$ and likewise
$H_p:T_pX\to\bigwedge^2(T_p^*X)$ for $p\in X$. This equation ties the
characteristic class of the $B$-field to the curvature of the
spacetime $X$. The {\it semi-classical limit} is the one in which $X$
``approaches'' flat space, i.e. $R(G)\to0$, and $B$ becomes
topologically trivial.

The important point is that both open and closed strings feel the
presence of $H$, but in very different ways. Closed strings only
``see'' the cohomology class $[H]$. According to the supergravity
equation (\ref{SUGRAeqs}), $[H]\to0$ in the semi-classical
limit. Thus a consistent semi-classical treatment of closed strings
will be insensitive to the presence of a $B$-field. Aspects of the
noncommutative geometry of closed strings can be found
in~\cite{CF1,FG1,FGR1,LSz1,LSz2,LLSz1}. They will not be covered in
these notes. In contrast, open
strings are sensitive to a concrete choice of $B$ with $H=\dd
B$. The induced two-form $x^*(B)\in\bigwedge^2(T^*\Sigma)$ does not
vanish in the limit $R(G)\to0$ and we can now explore the possibility
of analysing the string geometry semi-classically in the background
$B$-field, whereby one should be able to say some concrete things.

Consider a D-brane with embedding $\zeta:W\to X$. By a slight abuse of
notation, we will denote the pullback $\zeta^*(B)$ of the $B$-field
also by the symbol $B$, as this shouldn't cause any confusion in the
following. If this pullback is non-degenerate, then we can define the
{\it Seiberg-Witten bivector}~\cite{SW1}
\beq
\theta=\left(B+G\,B^{-1}\,G\right)^{-1} \ ,
\label{SWbivec}\eeq
which is regarded locally as a non-degenerate skew-symmetric linear
map $\theta_p:T_p^*X\to T_pX$ for $p\in X$.
In the formal ``limit'' $B\to\infty$, the Seiberg-Witten bivector is
$\theta=B^{-1}$. Thus when in addition $H=\dd B=0$, as happens in the
semi-classical flat space limit, the $B$-field defines a symplectic
structure on $W$ and $\theta$ is the Poisson bivector corresponding to the
symplectic two-form $B$. If $B$ is degenerate (and hence {\it not}
symplectic), then under favourable circumstances $\theta$ will still
be a Poisson bivector. One only requires that the Jacobi identity for
the corresponding Poisson brackets be fulfilled. This is equivalent to
the closure condition $\dd B=0$ only when $B$ is non-degenerate. The
quantum theory of the open strings attached to the D-brane tells us
that we should {\it quantize} this Poisson
geometry~\cite{AAS-J1,Chu-Ho1,Schom1,SW1}. This leads to a
string theoretic picture of the Kontsevich deformation quantization of
Poisson manifolds~\cite{Kont1}. An explicit realization of this
picture is provided by the Cattaneo-Felder topological
$\sigma$-model~\cite{CattFel1}.

When $\dd B\neq0$ new phenomena occur. One encounters generalizations
of ordinary Poisson structures, variations of quantum group algebras,
and the like~\cite{ARS1}. The generic situation leads to non-associative
deformations, which in some instances can still be handled by the
realization that they define $A_\infty$-homotopy associative
structures~\cite{CornSch1}. But there is no general notion of
quantization for such geometries. We will therefore continue to work
with the limits described above wherein one obtains true symplectic
geometries on the worldvolumes of D-branes. This sequence of limits is
often referred to as the {\it Seiberg-Witten limit}~\cite{SW1}.

A D-brane, and more generally collections of several D-branes, also
has much more structure to it than what we have described thus
far. The most primitive definition we can take of a D-brane is as a
{\it Baum-Douglas K-cycle} $(W,E,\zeta)$ in $X$~\cite{BD1}, where $W$ is a
spin$^c$ manifold, $\zeta:W\to X$ is a continuous map and $E\to W$ is a
complex vector bundle called the {\it Chan-Paton bundle}. The
description of D-branes in terms of K-cycles and
K-homology~\cite{AST1,HM1,RSz1,Sz1} will be
central to our analysis later on. If we equip $E$ with a smooth
connection, then we can define a Yang-Mills gauge theory on $W$. We
may also define more general field theories on $W$ by considering
smooth sections of this bundle (and other canonically defined bundles
over $W$). The connections and sections in this context are referred to
as {\it worldvolume fields}. The semi-classical motion of the D-brane
is thereby described dynamically by a {\it worldvolume field theory} which is
induced by the quantum theory of open strings on the brane. If the
geometry of the D-brane is quantized in the manner explained above,
then one finds a {\it noncommutative field theory} on the brane
worldvolume $W$ in the Seiberg-Witten limit~\cite{SW1}. In the
particular case of Yang-Mills theory, the deformation gives rise to a
{\it noncommutative gauge theory}.

In what follows, for us the interesting aspects of these
noncommutative worldvolume field theories will lie in the property
that they possess novel classical solitonic solutions which have no
counterparts in ordinary field theory~\cite{NCsol1}. In many
instances these solutions can themselves be interpreted as
D-branes~\cite{AGMS1,DMR1,HKL1,HKLM1,Witten1,GrossNek1,LPSz1,PopSz1,Sen1},
quite unlike the usual worldvolume field theories for $B=0$. Field
theoretical constructions of BPS soliton solutions in noncommutative
supersymmetric Yang-Mills theory, and their applications to D-brane
dynamics, can be found in~\cite{GrossNek1,Ohta1,Ohta2,Wimmer1}
(see~\cite{Ghosh1,Ghosh2} for BPS soliton solutions in other
noncommutative field theories). The
crucial point is that the Seiberg-Witten limit still retains a lot of
stringy information, in contrast to the usual field theoretic or point
particle limits of string theory. We can thus use the solitons of
noncommutative field theory to teach us about aspects of D-branes. For
example, they can provide insights into what sort of
worldvolume geometries live in a given spacetime $X$. Moreover, their
eventual quantization (which will not be covered here) could teach us
a lot about the nonperturbative structure of quantum string
theory. The purpose of these lecture notes is to explain this
correspondence in some specific spacetimes $X$, and to illustrate how
the techniques of noncommutative geometry can be used to construct the
appropriate noncommutative field configurations. The classical
noncommutative solitons will then admit an interpretation in terms of
``branes within branes''~\cite{Douglas1} and are built solely from the
properties of noncommutative field theory. A key point in our analysis
will be the unveiling of the connections with the K-theory
classification of D-brane
charges~\cite{MinMoore1,WittenK,HoravaK,OSz1}.

The outline of the remainder of these notes is as follows. In
Section~\ref{EuclD} we describe D-branes in flat Euclidean space and
the corresponding noncommutative field theories. This section
contains most of the elementary definitions used throughout these
notes. In Section~\ref{Solitons} we construct scalar field solitons on
Moyal spaces in these settings, and give their
interpretations as D-branes through an intimate connection to
K-homology. In Section~\ref{Gauge} we proceed to noncommutative gauge
theory and explicitly construct instanton solutions in the
two-dimensional case. In Section~\ref{TorD} we look at D-branes whose
worldvolume is a noncommutative torus and compare the instantons in
this case with those of the Moyal space. Finally, in
Section~\ref{GroupD} we consider D-branes in curved
spaces described by group manifolds and the ensuing fuzzy spaces
which describe the quantized worldvolume geometries, looking in
particular at the classic example of fuzzy spheres in $X={\rm
  SU}(2)\cong\Sphere^3$.

\newsection{Euclidean D-branes\label{EuclD}}

We will spend most of our initial investigation working in the
simplest cases of flat target spaces with $R(G)=H=0$. Then we are
automatically in the semi-classical regime of the string theory and we
can proceed straightforwardly with the quantization of the brane
worldvolume geometries as explained in Section~\ref{Intro}. In this
section we begin with some elementary definitions and then proceed to
describe the construction of noncommutative field theories on the
quantized worldvolumes. Extensive reviews of these sorts of
noncommutative field theories with exhaustive lists of references can
be found in~\cite{DougNekrev,KSrev,Szrev}.

\subsection{Moyal spaces\label{Defs}}

We consider spacetimes which are either a $d$-dimensional Euclidean space
$X=\real^d$ or a $d$-dimensional torus $X=\Torus^d$ for some integer
$d\geq2$. Let $2n\leq d$, and consider a D-brane localized along a
$2n$-dimensional hyperplane $\plane_{2n}\subset X$ with
tangent space $T\plane_{2n}$. Let $\theta:T^*\plane_{2n}\to T\plane_{2n}$ be a
skew-symmetric non-degenerate linear form. It may be represented by a
skew-symmetric $2n\times2n$ constant real matrix
$\theta=(\theta^{ij})_{1\leq i,j\leq2n}$ of maximal rank $2n$. There
exists a linear transformation on $\plane_{2n}\to \plane_{2n}$ which brings
$\theta$ into its canonical Jordan normal form
\beq
\theta=\left({}^{~\,0}_{-\theta_1}\,{}^{\theta_1}_{\,0}\right)
\oplus\cdots\oplus\left({}^{~\,0}_{-\theta_n}\,{}^{\theta_n}_{\,0}
\right) \ .
\label{thetanormal}\eeq
For definiteness we will take $\theta_k>0$ for all
$k=1,\dots,n$. It is straightforward to extend all of our
considerations below to matrices $\theta$ of rank~$<2n$ (and in
particular to odd-dimensional hyperplanes), but for simplicity we will
work only with non-degenerate $\theta$.

We now define the appropriate quantization of the D-brane worldvolume
$\plane_{2n}$ in these instances. Throughout we will denote
$\ii:=\sqrt{-1}$.
\begin{definition}
Let $F(\plane_{2n})=\complex\langle\one,x^1,\dots,x^{2n}\rangle$ be the
free unital algebra on $2n$ generators $x^1,\dots,x^{2n}$. Let
$I_\theta(\plane_{2n})$ be the two-sided ideal of $F(\plane_{2n})$ generated by
the $n\,(2n-1)$ elements $x^i\,x^j-x^j\,x^i-\ii\theta^{ij}\,\one$, $1\leq
i<j\leq2n$. The {\it Moyal $2n$-space} $\plane_{2n}^\theta$ is
the polynomial algebra
$\plane_{2n}^\theta=F(\plane_{2n})\,/\,I_\theta(\plane_{2n})$.
\label{Moyaldef}\end{definition}
\noindent
The Moyal $2n$-space will be loosely regarded as the algebra of
``functions generated by the coordinates'' $x^1,\dots,x^{2n}$
satisfying the commutation relations of a degree~$n$ Heisenberg
algebra
\beq
\left[x^i\,,\,x^j\right]:=x^i\,x^j-x^j\,x^i=\ii\theta^{ij}\,\one \ .
\label{Heisenalg}\eeq
In the normal form (\ref{thetanormal}), the only non-vanishing
commutation relations are
\beq
\bigl[x^{2k-1}\,,\,x^{2k}\bigr]=\ii\theta_k\,\one \ ,
\quad k=1,\dots,n \ .
\label{Heisennormal}\eeq
We will also formally regard $\plane_{2n}^\theta$ as a completion of the
polynomial algebra of Definition~\ref{Moyaldef}. There are various
technical complications with this since the polynomial algebra is
really an algebra of differential operators and so it has no
completion. We will not concern ourselves with this issue. Our
definition is such that the ``commutative limit''
$\plane_{2n}^{\theta=0}=\mathcal{S}(\plane_{2n})$ is the algebra of Schwartz
functions on $\plane_{2n}\to\complex$, i.e. the algebra of infinitely
differentiable functions whose derivatives all vanish at infinity
faster than any Laurent polynomial. This definition can be made more
precise through the formalism of deformation quantization which will
be considered in Section~\ref{Defquant}.

\subsection{Fock modules\label{Fock}}

To do concrete calculations later on, and in particular to define gauge
theories, we need to look at representations of the algebra
$\plane_{2n}^\theta$. For the moment, we set
$n=1$ and $\theta:=\theta_1>0$. In higher dimensions we can then
``glue'' the $n$ independent skew-blocks in (\ref{thetanormal})
together, as we explain later on. Starting from
the generators $x^1,x^2$ we introduce the formal complex linear
combinations
\beq
a=\mbox{$\frac1{\sqrt{2\,\theta}}$}\,\left(x^1+\ii x^2\right) \ ,
\quad a^\dag=\mbox{$\frac1{\sqrt{2\,\theta}}$}\,\left(x^1-\ii
  x^2\right) \ .
\label{creandef}\eeq
Then the Heisenberg commutation relation (\ref{Heisennormal}) is
equivalent to
\beq
\bigl[a\,,\,a^\dag\bigr]=\one \ .
\label{Heisenequiv1}\eeq

Consider the separable Hilbert space $\mathcal{F}:=\ell^2(\nat_0)$
with orthonormal basis $e_n$, $n\in\nat_0$. The dual space
$\mathcal{F}^*$ has corresponding basis $e_n^*$ with the canonical
dual pairing
\beq
\bigl\langle e_n^*\,,\,e_m\bigr\rangle=\delta_{nm} \ .
\label{dualpairing}\eeq
Operators from $\Fock\to\Fock$ are elements of the endomorphism algebra
${\rm End}(\Fock)\cong\Fock\otimes\Fock^*$.
\begin{definition}
The Hilbert space $\Fock$ is a finitely generated left
$\plane_2^\theta$-module, called the {\it Fock module}, with left
action $\plane_{2}^\theta\times\Fock\to\Fock$ given by
$$
a\cdot e_n=\sqrt{n}~e_{n-1} \ , \quad a^\dag\cdot e_n=
\sqrt{n+1}~e_{n+1} \ .
$$
\label{V2actiondef}\end{definition}
\noindent
To ease notation, we will not distinguish between the algebra
$\plane_2^\theta$ and its representation as operators in ${\rm End}(\Fock)$
acting on $\Fock$.
\begin{remark}
By the Stone-von~Neumann theorem, $\Fock$ is the unique irreducible
representation of the Heisenberg commutation relations. There is a
natural isomorphism $\Fock\cong{\rm L}^2(\real,\dd x)$ obtained by
representing $x^2=:M_x$ as multiplication of a function by $x\in\real$
and $x^1$ as the differential operator $\ii\theta\,\frac\dd{\dd
  x}$. This is known as the {\it Schr\"odinger representation}.
\label{Schrorem}\end{remark}

With this action one has
\bea
a\cdot e_0&=&0 \ , \nonumber\\[4pt]
\bigl(a^\dag\,a\bigr)\cdot e_n&=&n~e_n \ , \nonumber\\[4pt]
e_n&=&\mbox{$\frac1{\sqrt{n!}}$}\,\big(a^\dag\bigr)^n\cdot
e_0 \ .
\label{V2actionhas}\eea
Thus the orthonormal basis $e_n$ forms the complete set of eigenvectors for the
action of the element $a^\dag\,a\in \plane_2^\theta$ and is called
the {\it number basis}. The last identity in (\ref{V2actionhas})
implies that this is the natural basis for a given representation of
$a^\dag$ on some fixed vector $e_0\in\Fock$. For definiteness we will mostly
present our computations in the number basis, but it is possible to
reformulate everything in a basis independent way.
\begin{proposition}
The Fock module $\Fock$ is projective.
\label{Fockproj}\end{proposition}
\begin{proof}
The endomorphism
$$
\Pi_n:=e_n\otimes e_n^*
$$
is the orthogonal projection of $\Fock$ onto the one-dimensional
subspace spanned by the vector $e_n\in\Fock$. The
operators $\Pi_n$, $n\in\nat_0$ generate a complete system of mutually
orthogonal projectors with
$$
\Pi_n\,\Pi_m=\delta_{nm}~\Pi_n \ , \quad \sum_{n\in\nat_0}\,
\Pi_n=\one \ .
$$
For each $n\in\nat_0$, this determines an orthogonal decomposition for
the left action of the algebra $\plane_2^\theta$ on $\Fock$ given by
$$
\plane_2^\theta=\Pi_n\cdot \plane_2^\theta~\oplus~(\one-\Pi_n)\cdot
\plane_2^\theta \ .
$$
There is a natural isomorphism $\Fock\cong\Pi_n\cdot \plane_2^\theta$ given
by $\Pi_n\cdot f\mapsto f\cdot e_n\in\Fock$ for $f\in
\plane_2^\theta$. Thus $\Fock$ is projective.
\end{proof}
\begin{remark}
By formally iterating the orthogonal decomposition above we have
$$
\plane_2^\theta=\bigoplus_{n\in\nat_0}\,\Pi_n\cdot \plane_2^\theta
$$
with $\Pi_n\cdot \plane_2^\theta\cong\Fock$ for each
$n\in\nat_0$. Heuristically, this means that the Fock module is the
analog of a single ``point'' on the noncommutative space
$\plane_2^\theta$. This is analogous to what occurs in the commutative
situation, whereby any function $f$ can be decomposed formally as
$f(x)=\int\dd y~\delta(x-y)\,f(y)$ with the evaluation maps
$\delta_x(f)=f(x)$ being the analogs of the projectors $\Pi_n$ above.
\label{Fockpointrem}\end{remark}

Finally, the generalization to higher dimensions $n>1$ is obtained by
defining
\beq
a_k=\mbox{$\frac1{\sqrt{2\,\theta_k}}$}\,\bigl(x^{2k-1}+\ii x^{2k}
\bigr) \ , \quad a_k^\dag=\mbox{$\frac1{\sqrt{2\,\theta_k}}$}\,
\bigl(x^{2k-1}-\ii x^{2k}\bigr)
\label{creandimn}\eeq
for each $k=1,\dots,n$. Then the non-vanishing commutation relations
are given by
\beq
\bigl[a_k\,,\,a_l^\dag\bigr]=\delta_{kl}\,\one \ .
\label{Heisenequivn}\eeq
In this case a right module over the algebra $\plane_{2n}^\theta$ is obtained
by taking $n$ independent copies of the basic Fock module above with
$\Fock^{\oplus n}\cong\ell^2(\nat_0^n)\cong{\rm L}^2(\real^n,\dd
x)$. By the Hilbert hotel argument there is a natural isomorphism
$\Fock^{\oplus n}\cong\Fock$.
\begin{remark}
The commutation relations (\ref{Heisenequivn}) show that all algebras
$\plane_{2n}^\theta$ for $\theta_k\neq0$ are isomorphic and one can simply
scale away the noncommutativity parameters $\theta_k$ to~$1$, as the
redefinition (\ref{creandimn}) essentially does. At times it will be
convenient to keep $\theta$ in as an explicit parameter for comparison
of the new phenomena in noncommutative field theories to ordinary
field theories.
\label{Equivrem}\end{remark}

\subsection{Deformation quantization\label{Defquant}}

The algebra $\plane_{2n}^\theta$ can be regarded as a deformation
quantization of the algebra of Schwartz functions $\mcS(\plane_{2n})$ on
the {\it ordinary} hyperplane $\plane_{2n}$ in the standard
way~\cite{Bayen1} with
respect to the constant symplectic two-form $\omega=\theta^{-1}$. This
is common practise in the string theory literature. Although it will
not be used extensively in what follows, we will briefly describe here
the basic features of the approach. For most of these notes we will
stick to the more abstract setting with $\plane_{2n}^\theta$ realized as
operators on the Fock module $\Fock$ (or any other module), as
everything can be straightforwardly constructed in this setting. The
deformation quantization approach will only be used occasionally
when it can provide a useful way of envisaging the ``profiles'' of
noncommutative fields.

Consider a polynomial function $f:\plane_{2n}\to\complex$ with Fourier
transform $\tilde f:T\plane_{2n}\to\complex$ defined through
\beq
f(x)=(2\pi)^{-2n}\,\int_{T\plane_{2n}}\dd k~\tilde f(k)~
\exp\bigl(\ii\langle k,x\rangle\bigr) \ ,
\label{fFourierdef}\eeq
where $\langle k,x\rangle:=\sum_{i=1}^{2n}\,k_i\,x^i$ with
$k=(k_1,\dots,k_{2n})\in T\plane_{2n}$ and $x=(x^1,\dots,x^{2n})\in
\plane_{2n}$. Here we will distinguish between local coordinates $x^i$
on $\plane_{2n}$ and the generators $\hat x^i$ of the noncommutative algebra
$\plane_{2n}^\theta$ by drawing a hat over the latter symbols. Thus
$[\hat x^i,\hat x^j]=\ii\theta^{ij}\,\one$. We will usually consider
$\plane_{2n}^\theta$ in its concrete realization as linear operators in
${\rm End}(\Fock)$ acting on the Fock module, but this is not
necessary and the following construction also works at a more abstract
level~\cite{Szrev}.

The {\it Weyl symbol} $\Omega(f)\in \plane_{2n}^\theta$ is defined by
\beq
\Omega(f)=(2\pi)^{-2n}\,\int_{T\plane_{2n}}\dd k~\tilde
f(k)~\exp\bigl(\ii\langle k,\hat x\rangle\bigr) \ .
\label{Weyldef}\eeq
Heuristically, the Weyl symbol of $f(x)$ is obtained by substituting
in the generators of $\plane_{2n}^\theta$ to associate to the function a
{\it noncommutative field}
$\Omega(f)=f(\hat x)$. This definition extends to Schwartz functions
$f\in\mcS(\plane_{2n})$, for which $\Omega(f)$ is a compact operator when
acting on $\Fock$. The compact operators form a dense domain
$\compact=\compact(\Fock)$ in the endomorphism algebra ${\rm
  End}(\Fock)$. The exponential function of algebra elements appearing in
eq.~(\ref{Weyldef}) is defined formally through its power series
expansion. It implies a particular ordering of the noncommutative
variables called {\it Weyl ordering}.

The map $f\mapsto\Omega(f)$ is invertible. Its inverse allows one to
go the other way and associate functions to noncommutative fields. The
{\it Wigner function} $\Omega^{-1}({O})\in\mcS(\plane_{2n})$
of an element ${O}\in\compact$ is given by
\beq
\Omega^{-1}({O})(x)=\bigl|{\rm Pf}(2\pi\,\theta)\bigr|^{-1}\,
\int_{T\plane_{2n}}\dd k~\exp\bigl(-\ii\langle k,x\rangle\bigr)~
\Tr\left[\,{O}~\exp\bigl(\ii\langle k,\hat x\rangle\bigr)
\right] \ ,
\label{Wignerdef}\eeq
where the symbol ${\rm Pf}$ denotes the
Pfaffian of an antisymmetric matrix and $\Tr$ is the trace defined on
$\compact$. It follows that the Weyl symbol determines a vector space
isomorphism between appropriate subspaces of functions on $\plane_{2n}$ and
dense domains of elements in $\plane_{2n}^\theta$ in an appropriate
Fr\'echet algebra topology. This point of view is also useful for
adding further structure to the algebra $\plane_{2n}^\theta$. For instance,
the {\it trace} $\Tr$ on $\plane_{2n}^\theta$ is defined for
${O}\in\compact$ by
\beq
\Tr({O})=\int_{\plane_{2n}}\dd x~\Omega^{-1}({O})(x) \ .
\label{Tracedef}\eeq

However, the linear mapping $f\mapsto\Omega(f)$ is {\it not} an
algebra isomorphism. This fact can be used to {\it deform} the
pointwise multiplication of functions on $\plane_{2n}$ and define the {\it
  Moyal star-product} by
\beq
f\star g=\Omega^{-1}\bigl(\Omega(f)\,\Omega(g)\bigr)
\label{fstargdef}\eeq
for $f,g\in\mcS(\plane_{2n})$. For the domains of functions we are
interested in, a convenient explicit expression for the star-product
is
\beq
(f\star g)(x)=(2\pi)^{-2n}\,\int_{T\plane_{2n}}\dd k~\int_{\plane_{2n}}\dd y~
f\bigl(x+\mbox{$\frac12$}\,\theta\,k\bigr)~g\bigl(x+y
\bigr)~\exp\bigl(\ii\langle k,y\rangle\bigr)
\label{fstargexpl}\eeq
where $(\theta\,k)^i:=\sum_{j=1}^{2n}\,\theta^{ij}\,k_j$. With this
representation the star-product of two Schwartz functions is again a
Schwartz function. There are other commonly used explicit expressions
for the star-product in the string theory literature, such as a
Fourier integral representation or a formal asymptotic expansion using
a bidifferential operator~\cite{Szrev}, but these formulas do not necessarily
return Schwartz functions. Let us conclude the present discussion with
a class of examples that will be relevant to our later constructions.
\begin{example}
For $x=(x^1,x^2)\in \plane_2$ with $|x|^2:=(x^1)^2+(x^2)^2$ one can
straightforwardly compute the basic Gaussian Wigner
function~\cite{NCsol1,Lang1,LSzZ1}
$$
\Omega^{-1}(e_0\otimes e_0^*)(x)=2\e^{-|x|^2/\theta} \ .
$$
More generally, for any $n>m$ one finds the Wigner functions
$$
\Omega^{-1}(e_n\otimes e_m^*)(r,\vartheta)=2\,(-1)^m\,\sqrt{\mbox{$
\frac{m!}{n!}$}}\,\left(\mbox{$\frac{2\,r^2}\theta$}\right)^{
\frac{n-m}2}~\e^{\ii(n-m)\,\vartheta}~\e^{-r^2/\theta}~L^{n-m}_m\bigl(
\mbox{$\frac{2\,r^2}\theta$}\bigr)
$$
where $(r,\vartheta)\in[0,\infty)\times[0,2\pi)$ are plane polar
coordinates on $\plane_2$ and
$$
L_k^j(t)=\sum_{l=0}^k\,(-1)^l\,{{k+j}\choose{k-l}}~\frac{t^l}{l!}
$$
for $j,k\in\nat_0$ are the associated Laguerre polynomials. It is an
instructive exercise to check, using the explicit integral
representation (\ref{fstargexpl}), that the star-products of these
functions obey the appropriate orthonormality relations required for
both (\ref{dualpairing}) and (\ref{fstargdef}) to hold.
\label{starprodex}\end{example}

\subsection{Derivations\label{DerivMoyal}}

The infinitesimal action of the translation group of $\plane_{2n}$ induces
automorphisms $\partial_i:\plane_{2n}^\theta\to \plane_{2n}^\theta$,
$i=1,\dots,2n$. On generators they are given by
\beq
\partial_i\bigl(x^j\bigr)=\delta_i{}^j \ .
\label{partialgens}\eeq
With this definition one can verify the Leibniz rule, so that the
automorphisms $\partial_i$, $i=1,\dots,2n$ define a set of {\it
  derivations} of the algebra $\plane_{2n}^\theta$. One also finds the
expected commutation relations of the translation group
\beq
\bigl[\partial_i\,,\,\partial_j\bigr]=0 \ .
\label{translgprep}\eeq
It is possible to modify these relations to give a representation
$[\partial_i,\partial_j]=-\ii\Phi_{ij}\,\one$ of the centrally
extended translation group of $\plane_{2n}$ without affecting any of our
later considerations~\cite{AMNSz1}, but we will stick to the setting of
eq.~(\ref{translgprep}) for simplicity.

By using the Heisenberg commutation relations (\ref{Heisenalg}), one
finds for the representation of $\plane_{2n}^\theta$ on the Fock module
$\Fock$ that the derivations can be represented as {\it inner}
automorphisms
\beq
\partial_i(f)=\ii\,\sum_{j=1}^{2n}\,\bigl(\theta^{-1}\bigr)_{ij}\,\bigl[
x^j\,,\,f\bigr]
\label{partialinner}\eeq
for $f\in \plane_{2n}^\theta$. Moreover, they induce the Weyl symbols of
ordinary coordinate derivatives $\partial F/\partial x^i$ of Schwartz
functions $F$ through
\beq
\Omega\bigl(\mbox{$\frac{\partial F}{\partial x^i}$}\bigr)=
\partial_i\bigl(\Omega(F)\bigr) \ .
\label{Weylderiv}\eeq
Finally, one can show that eq.~(\ref{Tracedef}) defines an {\it
  invariant} trace for the action of the translation group since
\beq
\Tr\bigl(\partial_i({O})\bigr)=0
\label{Traceinv}\eeq
for any ${O}\in\compact$, which is equivalent to the usual
formula for integration by parts of Schwartz functions on
$\plane_{2n}$.

We now have all the necessary ingredients to study a broad class of
field theories on the noncommutative space $\plane_{2n}^\theta$. Elements
of the noncommutative algebra provide noncommutative fields, the
invariant trace gives us an integral, and the derivations introduced
above yield derivatives. We begin this investigation of noncommutative
field theories on the Euclidean D-brane worldvolume $\plane_{2n}$ in the
next section.

\newsection{Solitons on ${\plane^\theta_{2n}}$\label{Solitons}}

In this section we will study some elementary noncommutative {\it
  scalar} field theories on the worldvolume $\plane_{2n}$. We will
construct two broad classes of noncommutative solitons and describe
the rich geometric structure of the corresponding moduli spaces. We
shall then demonstrate that these solutions naturally define elements
of analytic K-homology~\cite{Witten2,Matsuo1,HM1}, which leads into their
worldvolume interpretation as D-branes in Type~II string
theory. Some reviews of noncommutative solitons in the contexts
described here can be found in~\cite{Haman1,Harvey1,Schap1}.

\subsection{Projector solitons\label{Projsoliton}}

Let ${\rm u}(\plane_{2n}^\theta)$ be the Lie algebra of {\it
  Hermitean} elements in the noncommutative space, i.e. the set of
$\phi\in\plane_{2n}^\theta$ for which the corresponding endomorphisms
of the Fock module are Hermitean operators with respect to the
underlying Hilbert space structure of $\Fock$, or equivalently
the corresponding Wigner functions $\Omega^{-1}(\phi):\plane_{2n}\to\real$
are real-valued. Let $V:\real\to\real$ be a polynomial function. We
may then define an {\it action functional} $S:{\rm
  u}(\plane_{2n}^\theta)\to\real$ by
\bea
S(\phi)&:=&\Tr\Bigl(\,\frac12\,\sum_{i=1}^{2n}\,\partial_i(\phi)\,
\partial_i(\phi)+V(\phi)\Bigr) \nonumber\\[4pt] &=&
\Tr\Bigl(-\frac12\,\sum_{i,j,k=1}^{2n}\,\bigl(\theta^{-1}\bigr)_{ij}\,
\bigl(\theta^{-1}\bigr)_{ik}\,\bigl[x^j\,,\,\phi\bigr]\,
\bigl[x^k\,,\,\phi\bigr]+V(\phi)\Bigr) \ ,
\label{IIAactiondef}\eea
when this expression makes sense. An action of this form describes the
dynamics of D-branes in {\it Type~IIA} string theory.

We now apply the variational principle. Identify the
tangent space $T\plane_{2n}$ with the hyperplane $\plane_{2n}$ itself. For any
$\alpha\in{\rm u}(\plane^\theta_{2n})$ and $t\in\real$, we may then
compute the variation of the action functional (\ref{IIAactiondef}) by
using invariance of the trace to get
\beq
\frac\delta{\delta\phi}S(\phi):=\frac\dd{\dd t}S(\phi+t\,\alpha)
\Big|_{t=0}=-\sum_{i=1}^{2n}\,(\partial_i\circ\partial_i)(
\phi)+V'(\phi) \ .
\label{varaction}\eeq
Setting this equal to~$0$ thereby gives the {\it equation of motion}
\beq
V'(\phi)=\sum_{i=1}^{2n}\,(\partial_i\circ\partial_i)(\phi) \ .
\label{eomreal}\eeq
We are interested in special classes of solutions to these equations.
\begin{definition}
A {\it soliton} on $\plane_{2n}^\theta$ is a solution $\phi\in{\rm
  u}(\plane_{2n}^\theta)$ of the equation of motion (\ref{eomreal}) for
which the action functional $S(\phi)$ is well-defined and finite.
\label{solitondef}\end{definition}

Naively, it seems to be very simple to construct soliton
solutions to the equations (\ref{eomreal}). If $\lambda$ is a critical
point of the polynomial $V$, i.e. $V'(\lambda)=0$, then an obvious
solution is the {\it constant} solution
$\phi_0=\lambda\,\one$. However, this solution is not trace-class and
so the action $S(\phi_0)$ is is not well-defined. To obtain finite action
solutions, we will first look for {\it static solitons} having
$\partial_i(\phi)=0$. From the inner derivation property
(\ref{partialinner}) we see that such fields live in the center of the
algebra~$\plane_{2n}^\theta$. We will soon lift this requirement and show how
to extend the construction to the general case. The advantage of this
restriction is that the equation of motion (\ref{eomreal}) for static
fields takes the very simple form
\beq
V'(\phi)=0 \ .
\label{staticeom}\eeq
Since $V$ is a polynomial, it is easy to find finite action
solutions of this equation~\cite{NCsol1}.
\begin{theorem}
Let $\lambda_1,\dots,\lambda_p$ be the critical points of the
polynomial function $V(\lambda)$. Then to each collection
$\P_1,\dots,\P_p$ of mutually orthogonal finite rank projectors on the
Fock module $\Fock$ there bijectively corresponds a static soliton
$$
\phi_{\{\P_i\}}=\sum_{i=1}^p\,\lambda_i~\P_i
$$
in ${\rm u}(\plane_{2n}^\theta)$ of action
$$
S\bigl(\phi_{\{\P_i\}}\bigr)=\bigl|{\rm Pf}(2\pi\,\theta)\bigr|^{-1}\,
\sum_{i=1}^p\,V(\lambda_i)~\Tr(\P_i) \ .
$$
\end{theorem}
\begin{remark}
In these notes all projectors are assumed to be Hermitean. The soliton
solution corresponding to the collection of projectors
with $k:=\Tr(\P_i)>0$ for some $i$ and $\Tr(\P_j)=0$ for all $j\neq i$
is interpreted as $k$ non-interacting solitons sitting on top of each
other at the origin of $\plane_{2n}$.
\label{ksolrem}\end{remark}
\begin{example}
The simplest projector on $\Fock$ is $\P_{(1)}=e_0\otimes e_0^*$. The
corresponding Wigner function is the Gaussian field computed in
Example~\ref{starprodex}. The soliton is localized within a width
$\theta^{-1/2}$ around the origin of the hyperplane. Note that this
width formally goes to infinity in the commutative limit
$\theta\to0$ and the action becomes infinite. The field
``delocalizes'' and spreads out to the constant solution of infinite
action. Since $\P_{(1)}$ has rank~$1$, it describes a single soliton
at the origin. More generally, the projector
$$
\P_{(k)}=\sum_{n=0}^{k-1}\,e_n\otimes e_n^*
$$
has rank $k$ and describes $k$ solitons at the origin $x=0$. The
corresponding Wigner functions are given by combinations of Gaussian
fields and Laguerre polynomials as described in
Example~\ref{starprodex}.
\label{projsolitonex}\end{example}

\subsection{Soliton moduli spaces\label{Solitonmod}}

In D-brane physics one would like to understand what are the parameters
that label inequivalent configurations modulo symmetries. These
configurations live in a moduli space which determines the effective
worldvolume geometries and on which we can study the
effective dynamics of the branes. This is also a crucial ingredient
for the eventual quantization of the systems, which would require an
integration over the moduli space. Let us introduce for each
$k\in\nat$ the complex {\it Grassmannian}
\beq
{\rm Gr}(k,\Fock)={\rm U}(\Fock)\,/\,{\rm U}(k)\times{\rm U}(\Fock) \
,
\label{grass}\eeq
where ${\rm U}(\Fock)$ is the group of unitary endomorphisms of the
Hilbert space $\Fock$ and ${\rm U}(k)$ is the group of $k\times k$
unitary matrices.
\begin{proposition}
The moduli space ${\cal M}_k^0(\plane_{2n}^\theta)$ of static
$k$-solitons is an infinite-dimensional K\"ahler manifold isomorphic
to the Grassmannian
$$
{\cal M}_k^0\big(\plane_{2n}^\theta\big)~=~{\rm
  Gr}\big(k,\Fock\big) \ .
$$
\label{staticsolprop}\end{proposition}
\begin{proof}
For static fields, the action (\ref{IIAactiondef}) is invariant under
the unitary transformations
$$
\phi~\longmapsto~{\rm Ad}_U(\phi)
$$
where $U\in{\rm U}(\Fock)$. Any two projectors on $\Fock$ of the same
rank are homotopic under this action. A projector of rank $k$ has
image which is a
$k$-dimensional linear subspace $\mathcal{F}_k\subset\Fock$ and thus
specifies a point in the Grassmannian (\ref{grass}), where the first
unitary group ${\rm U}(\Fock)$ acts on the whole of
$\mathcal{F}$, the group ${\rm U}(k)$ acts on the finite-dimensional
subspace $\Fock_k$, and the last ${\rm U}(\Fock)$ factor acts on the
orthogonal complement $\Fock\ominus\Fock_k\cong\Fock$. Let $E_k$ be the
tautological hyperplane bundle over ${\rm Gr}(k,\Fock)$. The inner
product on $E_k$ induces a natural metric on the determinant line
bundle $\det(E_k)$. The curvature two-form of this line bundle is the
natural K\"ahler form on the Grassmannian.
\end{proof}

All solitons $\phi$ in the infinite-dimensional moduli space ${\cal
  M}_k^0(\plane_{2n}^\theta)$ have the same action $S(\phi)=|{\rm
  Pf}(2\pi\,\theta)|^{-1}\,k\,V(\lambda_i)$ (for some $i$). We can
obtain a finite-dimensional soliton moduli space by
``translating'' the solitons obtained above away from the origin of
$\plane_{2n}$~\cite{NCsolmod1}. Introduce a complex structure on
$\plane_{2n}$ with local complex coordinates $z^j=x^{2j}+\ii
x^{2j-1}$, $\bar z^{\bar j}=x^{2j}-\ii x^{2j-1}$ for
$j=1,\dots,n$. For each $z=(z^1,\dots,z^n)$ we define the {\it
  coherent state} $\xi(z)\in\Fock$ by
\beq
\xi(z)=\exp\bigl(\,\mbox{$\sum\limits_{j=1}^n$}\,z^j\,a_j^\dag\,\bigr)\cdot
e_0 \ ,
\label{cohstatedef}\eeq
where again the exponential operator is understood through its formal
power series expansion. These vectors diagonalize the operators $a_j$
with
\beq
a_j\cdot\xi(z)=z^j~\xi(z) \ , \quad j=1,\dots,n \ .
\label{akdiag}\eeq
For the $k$-soliton solution, we place the solitons at some chosen
points $z_{(0)},z_{(1)},\dots,z_{(k-1)}$ in $\plane_{2n}$ with
$z_{(i)}=(z_{(i)}^1,\dots,z_{(i)}^n)$. Let $\P_{\{z_{(i)}\}}$ be the
orthogonal projection onto the linear span of the corresponding
vectors $\xi(z_{(0)}),\xi(z_{(1)}),\dots,\xi(z_{(k-1)})$. Then
$\Tr(\P_{\{z_{(i)}\}})=k$. The corresponding soliton solution is
called a {\it separated soliton}.
\begin{remark}
One can compute the Wigner functions corresponding to these
projectors. For $n=1$ and $z_{(i)}$ all distinct, one
finds~\cite{NCsolmod1}
$$
\Omega^{-1}\bigl(\P_{\{z_{(i)}\}}\bigr)(w,\bar w)=2\,
\sum_{i,j=0}^{k-1}\,\exp\bigl(\,\mbox{$-
(\frac{\bar w}{\sqrt\theta}-\bar z_{(i)})\,(\frac w{\sqrt\theta}-
z_{(j)})$}\,\bigr)
$$
for $(w,\bar w)\in\plane_2$. This function has a natural
interpretation in terms of separated solitons.
\end{remark}

The operators $\P_{\{z_{(i)}\}}$ may be characterized as those
projectors $\P$ obeying the equations
\beq
(\one-\P)\,a_j\,\P=0 \ , \quad j=1,\dots,n \ ,
\label{inveqs}\eeq
or equivalently that the image ${\rm im}(\P)\subset\Fock$ is an
invariant subspace for the collection of operators $a_1,\dots,a_n$. As
we will now show, they define a finite-dimensional subspace ${\cal
  M}_k(\plane_{2n}^\theta)\subset{\rm Gr}(k,\Fock)$. Introduce the
{\it Hilbert scheme} ${\rm Hilb}^k(\plane_{2n})$ of $k$
points in $\plane_{2n}\cong\complex^n$ as the space of ideals $\cal I$
of codimension $k$ in the polynomial ring
$\complex[y^1,\dots,y^n]$. It is easy to see at a heuristic level how
the Hilbert scheme is related to the projectors constructed
above. Since $f\in{\cal I}$ implies that $f\,g\in{\cal I}$ for all
polynomial functions $g$, the polynomials in an ideal $\cal I$ may be
thought of roughly as projections on $\complex[y^1,\dots,y^n]\to{\cal
  I}$. Conversely, if $\P=\P_{\{z_{(i)}\}}$ with all $z_{(i)}$
distinct, then the corresponding ideal $\cal I$ consists of those
polynomials $f\in\complex[y^1,\dots,y^n]$ which vanish at the loci
$z_{(i)}$, i.e. $f(z_{(i)}^1,\dots,z_{(i)}^n)=0$ for each
$i=0,1,\dots,k-1$. This correspondence can be made more precise.
\begin{theorem}
The moduli space ${\cal M}_k(\plane_{2n}^\theta)$ of separated
$k$-solitons is the Hilbert scheme
$$
{\cal M}_k\big(\plane_{2n}^\theta\big)~=~{\rm Hilb}^k\big(\plane_{2n}
\big) \ .
$$
\label{sepmodspace}\end{theorem}
\begin{proof}
We set up a one-to-one correspondence between projectors $\P\in{\cal
  M}_k(\plane_{2n}^\theta)$ obeying eq.~(\ref{inveqs}) and ideals
${\cal I}\in{\rm Hilb}^k(\plane_{2n})$. Define for each
polynomial $f\in\complex[y^1,\dots,y^n]$ the vector
$$
e_f=f\big(a_1^\dag,\dots,a_n^\dag\big)\cdot e_0
$$
in $\Fock$. If ${\cal I}\subset\complex[y^1,\dots,y^n]$ is an ideal of
codimension $k$, then we let $\one-\P$ be the orthogonal projection of
$\Fock$ onto the linear span $\bigoplus_{f\in{\cal I}}\,\complex\cdot
e_f$. Conversely, if $\P\in{\cal M}_k(\plane_{2n}^\theta)$ we set
$${\cal I}=\bigl\{f\in\complex[y^1,\dots,y^n]~\big|~\P\cdot e_f=0\bigr\} \
, $$ which is an ideal since $\P\,a_j^\dag=\P\,a_j^\dag\,\P$ for each
$j=1,\dots,n$.
\end{proof}

\begin{example} Theorem~\ref{sepmodspace} allows us to work out some
  explicit soliton moduli spaces for low values of the integers $k$
  and $n$.
\begin{enumerate}
\item ${\cal M}_k(\plane_2^\theta)={\rm Hilb}^k(\plane_2)$ is the
  $k$-th symmetric product orbifold ${\rm Sym}^k(\plane_2)=(\plane_2)^k/S_k$,
  where the symmetric group $S_k$ acts on the soliton positions in
  $(\plane_2)^k$ by permuting the entries of a $k$-tuple of elements
  of the plane $\plane_2$. The K\"ahler metric inherited from the
  Grassmannian is smooth at the orbifold points corresponding to
  coinciding soliton positions in
  $(\plane_2)^k$~\cite{NCsolmod1,Sweden1}. Thus noncommutativity
  smooths out the orbifold singularities of the symmetric product and
  as K\"ahler manifolds one has an isomorphism
$$
{\cal M}_k\big(\plane_2^\theta\big)=\big(\plane_2\big)^k \ .
$$
\item The two-soliton moduli space is~\cite{NCsolmod1}
$$
{\cal M}_2\big(\plane_{2n}^\theta\big)=\plane_{2n}\times
{\cal O}_{\mathbb{P}^{n-1}}\big(-1\big)
$$
where ${\cal O}_{\mathbb{P}^{n-1}}(-1)\to\mathbb{P}^{n-1}$ is the Hopf
bundle over the complex projective space
$\mathbb{P}^{n-1}$. The first factor describes the center of mass
position of the soliton configuration, while the second factor is the
resolution of the singularity of the moduli space for the relative
position which blows up the origin into $\mathbb{P}^{n-2}$. Again the
moduli space is non-singular.
\item ${\cal M}_k(\plane_4^\theta)={\rm Hilb}^k(\plane_4)$ is a smooth
  manifold which is a crepant resolution of the singular quotient
  variety ${\rm Sym}^k(\plane_4)$. It also arises as the moduli space
  of $k$ ${\rm U}(1)$ instantons on
  $\plane_4^\theta$~\cite{BradenNek1,NCsolmod1}. However, while the
  instanton moduli
  space is endowed with a hyper-K\"ahler metric, the soliton moduli
  space ${\cal M}_k(\plane_4^\theta)$ is only a K\"ahler manifold.
\item For $n>2$ and $k>3$ the moduli space ${\cal
    M}_k(\plane_{2n}^\theta)$ generically contains branches of varying
  dimension and so is not even a manifold~\cite{JapScheme1}.
\end{enumerate}
\label{solmodex}\end{example}

\subsection{Partial isometry solitons\label{Partiso}}

We will now construct solitons corresponding to general complex
elements $\phi\in\plane_{2n}^\theta$. We call these {\it complex
solitons}. We use the same polynomial
function $V$ as before, but now the action functional
$S:\plane_{2n}^\theta\to\real$ is defined by
\beq
S\bigl(\phi\,,\,\phi^\dag\bigr)=
\Tr\Bigl(\,\sum_{i=1}^{2n}\,\partial_i\big(\phi\big)\,
\partial_i\big(\phi^\dag\big)+V\big(\phi^\dag\phi-\one\big)+
V\big(\phi\,\phi^\dag-\one\big)\,\Bigr) \ .
\label{actioncomplex}\eeq
Such an action describes the dynamics of D-branes in {\it Type~IIB}
string theory. As before it is straightforward to obtain soliton
solutions of the equations of motion corresponding to
(\ref{actioncomplex}).
\begin{proposition}
To each partially isometric Fredholm operator on the Fock module
$\Fock$ there bijectively corresponds a static complex soliton.
\label{partisosol}\end{proposition}
\begin{proof}
Varying $\phi$ and $\phi^\dag$ independently in the action functional
(\ref{actioncomplex}) shows that the equations of motion for static
fields are satisfied if $\phi,\phi^\dag$ obey the defining equation
$$
\phi\,\phi^\dag\,\phi=\phi
$$
for a partial isometry of $\Fock$. Equivalently, $\phi$ is an isometry
in the orthogonal complement to a kernel and cokernel, or
$$
\phi^\dag\,\phi=\one-\P_{\ker(\phi)} \ , \quad
\phi\,\phi^\dag=\one-\P_{{\rm coker}(\phi)}
$$
where $\P_{\ker(\phi)}$ and $\P_{{\rm coker}(\phi)}$ are the
orthogonal projections onto the kernel and cokernel of
$\phi$. Substituting these expressions into the action functional
(\ref{actioncomplex}) and demanding that it be finite requires that
both $\ker(\phi)$ and ${\rm coker}(\phi)$ be finite-dimensional
subspaces of $\Fock$, i.e. that $\phi$ be also a Fredholm operator.
\end{proof}
\begin{remark}
Using Remark~\ref{Fockpointrem}, the finite-dimensional subspaces
$\ker(\phi)$ and ${\rm coker}(\phi)$
are identified with the vanishing locus of the complex soliton
in the corresponding Wigner function formulation on $\plane_{2n}$.
\label{supportrem}\end{remark}
\begin{definition}
The {\it topological charge} $Q(\phi)$ of a complex
soliton $\phi\in\plane_{2n}^\theta$ is its analytic index
$Q(\phi):={\rm index}(\phi)=\dim\ker(\phi)-\dim\,{\rm coker}(\phi)$.
\label{topchargedef}\end{definition}

To explicitly construct such solitons, let ${\rm
  C}\ell(\plane_{2n})$ be the complex Clifford algebra of the vector
space $\plane_{2n}$ equipped with the canonical quadratic form. Let
$\Delta_{\pm}$ be the irreducible half-spinor modules over ${\rm
  C}\ell(\plane_{2n})$ of ranks~$r:=2^{n-1}$. The half-spinor
generators are denoted $\sigma_i:\Delta_+\to\Delta_-$, $i=1,\dots,2n$
and they satisfy the algebra
\beq
\sigma_i{}^\dag\,\sigma_j+\sigma_j{}^\dag\,\sigma_i~=~2\,
\delta_{ij}\,\one_{r}~=~\sigma_i\,\sigma_j{}^\dag+
\sigma_j\,\sigma_i{}^\dag \ .
\label{Cliffalg}\eeq
We introduce the operator
\beq
\sigma_x:=\sum_{i=1}^{2n}\,\sigma_i\otimes x^i
\label{sigmaxdef}\eeq
regarded as an element $\sigma_x\in{\rm
  Hom}(\Delta_+\otimes\Fock,\Delta_-\otimes\Fock)$.
\begin{lemma}
The operator $\sigma_x$ has one-dimensional kernel and no
cokernel.
\label{sigmaxlemma}\end{lemma}
\begin{proof}
{}From eqs.~(\ref{Heisenequivn}) and (\ref{Cliffalg}) it follows that
the operator (\ref{sigmaxdef}) and its adjoint obey the identities
\bea
\sigma_x\,\sigma_x{}^\dag&=&\sum_{i=1}^n\,\one_{r}\otimes2\,
\theta_i\,\bigl(a_i^\dag\,a_i+\mbox{$\frac12$}\,\one\bigr)-
\sum_{i,j=1}^{2n}\,\ii\theta^{ij}\,\Sigma_{ij}\otimes\one \ , \nonumber\\[4pt]
\sigma_x{}^\dag\,\sigma_x&=&\sum_{i=1}^n\,\one_{r}\otimes2\,
\theta_i\,\bigl(a_i^\dag\,a_i+\mbox{$\frac12$}\,\one\bigr)-
\sum_{i,j=1}^{2n}\,\ii\theta^{ij}\,\Sigma_{ij}{}^\dag\otimes\one
\nonumber\eea
where
$$
\Sigma_{ij}=\mbox{$\frac14$}\,\bigl(\sigma_i\,\sigma_j{}^\dag-\sigma_j\,
\sigma_i{}^\dag\bigr) \ , \quad
\Sigma_{ij}{}^\dag=\mbox{$\frac14$}\,\bigl(\sigma_i{}^\dag\,\sigma_j-
\sigma_j{}^\dag\,\sigma_i\bigr) \ .
$$
By elementary group theory, the last term in the second product is
diagonalized by the lowest weight spinor $\psi_0$ of ${\rm SO}(2n)$ to
$-\sum_{i=1}^{n}\,\theta_i\,\one_{r}\otimes\one$. Along with
eq.~(\ref{V2actionhas}), this
implies that the operator $\sigma_x$ has a
one-dimensional kernel in $\Delta_+\otimes\Fock$ which is spanned by
the vector $\psi_0\otimes e_0$. On the other hand, the right-hand side
of the first product can never vanish and so the kernel of
$\sigma_x{}^\dag$ is trivial.
\end{proof}
\begin{theorem}
The surjection ${\sf T}\in{\rm
  Hom}(\Delta_+\otimes\Fock,\Delta_-\otimes\Fock)$ defined by
$$
{\sf T}=\bigl(\sigma_x\,\sigma_x{}^\dag\bigr)^{-1/2}\,
\sigma_x
$$
is a complex soliton of topological charge~$Q(\T)=1$.
\label{tachyondef}\end{theorem}
\begin{proof}
By Lemma~\ref{sigmaxlemma} the positive operator
$\sigma_x\,\sigma_x{}^\dag$ is invertible and so the operator
${\sf T}$ is well-defined. Furthermore, it is a partial isometry,
$\T\,\T^\dag\,\T=\T$, and has one-dimensional kernel and trivial
cokernel with
$$
\T\,\T^\dag=\one_{r}\otimes\one \ , \quad \T^\dag\,\T=
\one_{r}\otimes\one-\P_{\ker\sigma_x}
$$
implying that ${\rm im}(\T)=\Delta_-\otimes\Fock$.
\end{proof}
\begin{remark}
The classical solution of Theorem~\ref{tachyondef} is interpreted as
a single-soliton solution. More generally, a $k$-soliton solution
is given by the power $(\T)^k$ which has no cokernel and
$k$-dimensional kernel by the Boutet~de~Monvel index theorem~\cite{BdeM1},
i.e. $Q(\T)^k=k$. One can also construct separated complex
solitons by ``translating'' these operators away from the origin to
points $x_{(0)},x_{(1)},\dots,x_{(k-1)}\in\plane_{2n}$ and
defining~\cite{MartMoore1}
$$
\T_{\{x_{(i)}\}}=\prod_{i=0}^{k-1}\,\bigl(\sigma_{x-x_{(i)}\,\id}\,
\sigma_{x-x_{(i)}\,\id}{}^\dag\bigr)^{-1/2}\,\sigma_{x-x_{(i)}\,\id} \ .
$$
The analysis of the moduli space of these separated complex solitons
is similar to that carried out in Section~\ref{Solitonmod}.
\label{sepcomplexrem}\end{remark}

\subsection{Topological charges\label{Topchargessec}}

We will now derive a geometric formula for the topological charge of a
complex soliton by relating the analytic index of $\T$ to a
topological index~\cite{HM1,Matsuo1}. Let
$\Sphere^{2n-1}=\{x\in\plane_{2n}~|~|x|=1\}$ be the unit sphere of odd
dimension $2n-1$ in the hyperplane $\plane_{2n}$. The restriction map
$\plane_{2n}\,\setminus\,\{0\}\to\Sphere^{2n-1}$ is defined by $x\mapsto \frac
x{|x|}$. The soliton $\T$ constructed in Theorem~\ref{tachyondef} can
be thought of as a noncommutative version of the map
$\mu:\Sphere^{2n-1}\to{\rm GL}(r,\complex)$ defined by
\beq
\mu_x=\sum_{i=1}^{2n}\,\frac{x^i}{|x|}\,\sigma_i \ ,
\label{betaxdef}\eeq
which is Clifford multiplication on the ${\rm
  C}\ell(\plane_{2n})$-module $\Delta_+$ by the vector $x\in\plane_{2n}$.
To make this correspondence more precise, we have to explain what we
mean by restricting an operator to a noncommutative sphere.

We begin by choosing another polarization of the Fock module $\Fock$
which can be regarded as the holomorphic version of the Schr\"odinger
representation defined in Remark~\ref{Schrorem}. The {\it Bargmann
  quantization} is the natural isomorphism
\beq
\Fock\cong\Fock_{\rm B}:={\rm Hol}\bigl(\plane_{2n}\,,\,\exp(-2\,
\mbox{$\sum\limits_{k=1}^n$}\,\theta_k\,|z^k|^2)~\dd z\bigr)
\label{bargmanndef}\eeq
of $\plane_{2n}^\theta$-modules obtained by representing $a_k=M_{z^k}$
as multiplication by the
complex coordinate $z^k$ and $a_k^\dag$ as the partial differential
operator $-\frac\partial{\partial z^k}$. In Bargmann quantization, an
orthogonal basis is provided by the collection of monomials
\beq
z^{\vec k}:=\prod_{i=1}^n\,\big(z^i\big)^{k_i} \ , \quad
\vec k=(k_1,\dots,k_n)\in\nat_0^n \ .
\label{Bargmannbasis}\eeq

The advantage of the diffeomorphism (\ref{bargmanndef}) is that there is
a precise way to restrict vectors in $\Fock_{\rm B}$ to the sphere
$\Sphere^{2n-1}$. Using polar coordinates we decompose the canonical
measure on $\plane_{2n}$ as $\dd z=r^{2n-1}~\dd r~\dd\Omega$, where
$r\in\real_+:=[0,\infty)$ and $\dd\Omega$ is the
standard round measure on the unit sphere $\S^{2n-1}$. We use
homogeneous complex coordinates $(z,\bar z)$ on $\S^{2n-1}$ with
$\sum_{k=1}^n\,|z^k|^2=1$.
The {\it Hardy space} ${\rm H}^2(\S^{2n-1},\dd\Omega)$ is the closed Hilbert
subspace of ${\rm L}^2(\S^{2n-1},\dd\Omega)$ consisting of ${\rm
  L}^2$-functions on $\S^{2n-1}$ which admit a holomorphic extension
to all of $\plane_{2n}$. An orthogonal basis for the Hardy space ${\rm
  H}^2(\S^{2n-1},\dd\Omega)$ is again provided by the
monomials~\cite{HM1}
\beq
\varphi_{\vec k}:=z^{\vec k} \ .
\label{Hardybasis}\eeq

To specify the restriction of the $\plane_{2n}^\theta$-module
structure, we need to make sense of the action of the classical
coordinates $z^k,\bar z^{\bar k}$ on ${\rm H}^2(\S^{2n-1},\dd\Omega)$.
Let $f:\S^{2n-1}\to\complex$ be an ${\rm L}^2$-function. Let $P_+:{\rm
  L}^2(\S^{2n-1},\dd\Omega)\to{\rm H}^2(\S^{2n-1},\dd\Omega)$ be the
{\it Szeg\'o projection} defined by
\beq
(P_+f)(z)=\int_{\S^{2n-1}}\,\dd\Omega(w,\bar w)~\frac{f(w,\bar w)}
{\Bigl(1-\sum\limits_{k=1}^n\,z^k\,\bar w^{\bar k}\Bigr)^n} \ ,
\label{Szegoproj}\eeq
and let $M_f:{\rm H}^2(\S^{2n-1},\dd\Omega)\to{\rm
  L}^2(\S^{2n-1},\dd\Omega)$ be the operator of multiplication by
$f$. The {\it Toeplitz operator} $\T_f:{\rm
  H}^2(\S^{2n-1},\dd\Omega)\to{\rm H}^2(\S^{2n-1},\dd\Omega)$ is
then defined by
\beq
\T_f=P_+\circ M_f \ .
\label{Toeplitzdef}\eeq
It is the {\it compression} of the multiplication operator $M_f$ to
${\rm H}^2(\S^{2n-1},\dd\Omega)$. The action of the Toeplitz operators
corresponding to the coordinate
functions on the basis (\ref{Hardybasis}) is straightforward to work
out. Let $\vec e_i$ denote the $i$-th standard unit vector in
$\real^n$. Then for each $i=1,\dots,n$ one has
\bea
\T_{z^i}\varphi_{\vec k}&=&\varphi_{\vec k+\vec e_i} \ ,
\nonumber\\[4pt] \T_{\bar z^{\bar\imath}}\varphi_{\vec k}&=&\left\{
\begin{matrix}0&\quad,\quad&\quad k_i=0 \\ 2\pi\,\frac{k_i}{|\vec k|
+n-1}\,\varphi_{\vec k-\vec e_i} &\quad,\quad&\quad k_i>0
\end{matrix} \right.
\label{Toeplitzcoords}\eea
where $|\vec k|:=\sum_{i=1}^n\,k_i$ for $\vec k\in\nat_0^n$.
\begin{example}
It is instructive at this stage to look explicitly at the
two-dimensional case $n=1$. Then $r=1$, $\sigma_1=1$ and
$\sigma_2=\ii$ so that the standard complex soliton
$$
\T=\bigl(a^\dag\,a\bigr)^{-1/2}\,a=:\shift^\dag
$$
coincides with the {\it shift operator} $\shift:\Fock\to\Fock$ defined
on the number basis by
$$
\shift e_m=e_{m+1} \ .
$$
The shift operator is the basic partial isometry of $\Fock$ with
$$
\shift^\dag\,\shift=\one \ , \quad \shift\,\shift^\dag=
\one-e_0\otimes e_0^* \ ,
$$
and hence it has no kernel and a one-dimensional cokernel spanned by
the lowest vector $e_0$ in the number basis of $\Fock$. The Hardy
space in this instance ${\rm H}^2(\S^1,\dd\Omega)=\Hil_+(\S^1)$ is the
closed Hilbert subspace of
${\rm L}^2(\S^1,\dd\Omega)$ spanned by the non-negative Fourier modes on
the circle $\varphi_k=\e^{\ii k\,\Omega}$, $k\geq0$,
$\Omega\in[0,2\pi)$. This is the positive eigenspace of the Dirac
operator $-\ii\frac\dd{\dd\Omega}$ on $\S^1$. By identifying the
monomial $\varphi_k$ with the number basis element $e_k\in\Fock$, one
can identify the Toeplitz operators $\T_{\varphi_k}=(\shift)^k$ for
$k>0$ (which have non-trivial kernels for $k<0$).
\label{Toeplitzex}\end{example}

The correspondence between Toeplitz operators and complex solitons on
$\plane_{2n}^\theta$ given in Example~\ref{Toeplitzex} can be
generalized to higher dimensions $n>1$ by replacing the Hardy space with
${\rm H}^2(\S^{2n-1},\dd\Omega)\otimes\complex^r$ (with $r:=2^{n-1}$
as before) and extending the Toeplitz operators $\T_f$ to matrix-valued ${\rm
  L}^2$-functions $f:\S^{2n-1}\to\mat_r(\complex)$. Then the complex
soliton of Theorem~\ref{tachyondef} corresponds to a Toeplitz operator
\beq
\T_\mu=P_+\circ M_\mu
\label{Toeplitzsoliton}\eeq
in ${\rm End}({\rm H}^2(\S^{2n-1},\dd\Omega)\otimes\Delta)$, where
$\Delta$ is the unique irreducible spinor module of rank~$r$ over the Clifford
algebra ${\rm C}\ell(\real^{2n-1})$. This defines a bounded Fredholm
operator on Hardy space and we finally arrive at our desired geometric
formula for the topological charge.
\begin{theorem}
The topological charge of the complex soliton $\T$ is given by the
characteristic class formula
$$
Q(\T)={\rm ch}(\mu)\big[\S^{2n-1}\big]
$$
where
$$
{\rm
  ch}(\mu)=\mu^*\bigl(\,\mbox{$\sum\limits_{j=1}^n\,\frac{(-1)^j}
{(j-1)!}\,\omega_{2j-1}$}\,\bigr)
$$
and $\omega_i$ are the standard generators of the rational cohomology
${\rm H}^i({\rm GL}(r,\complex),\rat)$.
\label{topchargethm}\end{theorem}
\begin{proof}
Represent the partial isometry $\T$ in Bargmann quantization. Then the
Toeplitz operator (\ref{Toeplitzsoliton}) is the image of $\T$ under
the restriction map $\Fock_{\rm B}\otimes\Delta\to{\rm
  H}^2(\S^{2n-1},\dd\Omega)\otimes\Delta$. This map is not unitary,
but it is a bijection and consequently ${\rm index}(\T)={\rm
  index}(\T_\mu)$. The result now follows from the Boutet~de~Monvel
index theorem~\cite{BdeM1} applied to the Toeplitz operator $\T_\mu$
and the fact that the sphere $\S^{2n-1}$ has trivial Todd class.
\end{proof}
\begin{remark}
For $n=1$, Theorem~\ref{topchargethm} reads
$Q(\T)=\frac1{2\pi\ii}\,\int_{\S^1}\,\mu^{-1}~\dd\mu$ and thus the
topological charge of the noncommutative soliton coincides with the
winding number of the function $\mu:\S^1\to\S^1$.
\label{windingrem}\end{remark}

\subsection{Worldvolume construction\label{Worldvolume}}

We will now demonstrate how the soliton solution constructed in
Section~\ref{Partiso} has a natural interpretation in terms of
D-branes. The construction of these solitons is intimately related to
the Atiyah-Bott-Shapiro (ABS) construction $M_\sharp\,{\rm
  Spin}(\plane_{2n})\to\K^\sharp(\plane_{2n})$ of K-theory classes in terms
of Clifford modules, whose generator is provided by Clifford
multiplication (\ref{betaxdef}). By Theorem~\ref{topchargethm}, the
topological charge of the noncommutative soliton coincides with the
index of the classical ABS class of $\mu$, whose winding number
determines {\it D-brane charge}~\cite{WittenK,HoravaK,OSz1}. By
elucidating this point we will link our solutions naturally with
D-branes. Our ensuing worldvolume
interpretation will thereby demonstrate the equivalence between the
usual commutative and the noncommutative descriptions of D-branes, and
will further provide a novel insight into the nature of the
worldvolume geometries.

The basic idea behind the construction is to associate D-branes
to algebras of ``almost commuting'' operators~\cite{HM1}. Let
$\lin(\Fock)\subset{\rm End}(\Fock)$ be
the $C^*$-algebra of bounded linear operators on the Hilbert space
$\Fock$. The $C^*$-algebra $\compact=\compact(\Fock)$ of compact
operators on $\Fock$ is a closed ideal in $\lin(\Fock)$. The Toeplitz
operators (\ref{Toeplitzdef}) generate a unique $C^*$-algebra called the
{\it Toeplitz algebra} which we will denote by
$\alg\subset\lin(\Fock)$. To ease notation we write
$X:=\S^{2n-1}$. Let $C(X)$ be the commutative $C^*$-algebra of continuous
complex-valued functions on $X$. In general, the map $C(X)\to\alg$
defined by $f\mapsto\T_f$ is not an algebra homomorphism.
\begin{proposition}
For each pair of functions $f,g\in C(X)$, the difference
$\T_f\,\T_g-\T_{f\,g}$ in the Toeplitz algebra $\alg$ is a
compact operator on the Fock module $\Fock$.
\label{Toeplitzcomplem}\end{proposition}

It follows that $[\T_f,\T_g]\in\compact$ is compact for any $f,g\in
C(X)$~\cite{BDF1}, and compact operators are always regarded as
``small'' (being elements of a closed dense domain in
$\lin(\Fock)$). Thus the Toeplitz algebra $\alg$ is ``almost
commuting''.  We can identify operators which differ from one another
only by a ``small'' perturbation by regarding them as elements of the
{\it Calkin algebra} $\Calkin(\Fock):=\lin(\Fock)\,/\,\compact$ with
the natural projection $\pi:\lin(\Fock)\to\Calkin(\Fock)$. The Calkin
algebra is a unital $C^*$-algebra. In this way our explicit
construction of static complex solitons in Section~\ref{Partiso} leads
naturally to the Brown-Douglas-Fillmore classification of
essentially normal operators~\cite{BDF1,HM1}.

Starting from the soliton configuration (\ref{Toeplitzsoliton}), the
map $\T_\mu\mapsto\mu$ gives rise to a $C^*$-epimorphism
$\beta:\alg\to C(X)\otimes\mat_r(\complex)$. The algebra
$C(X)\otimes\mat_r(\complex)$ is Morita equivalent to $C(X)$, and
hence it will suffice to restrict our attention to the commutative
algebra $C(X)$. It follows that the Toeplitz operators generate an
{\it extension} of the commutative algebra of functions on $X$ by compact
operators, i.e. the noncommutative algebra $\alg$ fits into
a short exact sequence
\beq
0~\longrightarrow~\compact~\longrightarrow~\alg~
\stackrel{\beta}{\longrightarrow}~C(X)~\longrightarrow~0 \ .
\label{extexactseq}\eeq
Exactness of the sequence follows from
Proposition~\ref{Toeplitzcomplem} and the fact that
$\T_f\,\T_g-\T_{f\,g}\in\ker(\beta)$ for any two functions $f,g\in
C(X)$.

We can introduce a set of equivalence classes of extensions
(\ref{extexactseq}) as follows. Define a map
$\tau:C(X)\to\Calkin(\Fock)$ called the {\it Busby invariant} by
$\tau(f)=\pi(\T_f)$. By Proposition~\ref{Toeplitzcomplem}, the Busby
invariant is a unital
$C^*$-monomorphism with $\alg=\pi^{-1}({\rm im}\,\tau)$. Any extension
(\ref{extexactseq}) can be uniquely characterized by a pair
$(\Hil,\tau)$, where $\Hil$ is a separable Hilbert space and
$\tau:C(X)\to\Calkin(\Hil)$ is a unital $C^*$-monomorphism~\cite{BDF1}. On the
collection of pairs $(\Hil,\tau)$, there is a natural notion of
unitary (or strong) equivalence and a natural direct sum
operation~\cite{BDF1,BD1,HM1}. The set $\Ext(C(X),\compact)$ of
equivalence classes of extensions (\ref{extexactseq}) is thus a
semigroup. An extension is {\it trivial} if the exact sequence
(\ref{extexactseq}) splits, i.e. if the corresponding Busby invariant
$\tau$ has a lift to all of $\lin(\Hil)$. The quotient of
$\Ext(C(X),\compact)$ by trivial extensions is an abelian group which
defines a dual homology theory to the K-theory of $X$. This is called
the {\it analytic K-homology group}~$\K_1^{\rm a}(X)$.

The constructions of this section bring us finally to
our main result~\cite{BD1}.
\begin{theorem}
There is a one-to-one correspondence between static complex solitons
$\T\in\plane_{2n}^\theta$ with Toeplitz extension classes
(\ref{extexactseq}) in $\K_1^{\rm a}(X)$ and D-branes $(W,E,\zeta)$ on
$X$ with odd-dimensional worldvolumes $W$.
\label{Dbranethm}\end{theorem}
\begin{proof}
Let $\Delta_W\to W$ be the spin$^c$ bundle over the odd-dimensional
spin$^c$ manifold $W$. Let
$\Hil={\rm L}^2(W,\Delta_W\otimes E)$ be the Hilbert space of
square-integrable sections of the twisted spin$^c$ bundle over the
worldvolume. Equip the Chan-Paton bundle $E\to W$ with a connection,
and let $W$ inherit the metric from $X$ by pullback under the continuous
map $\zeta:W\to X$. The corresponding
twisted Dirac operator $\Dirac_E$ can be viewed as a closed unbounded
operator $\Dirac_E:\hil\to\hil$. If the connection and metric are
generic, then $\Dirac_E$ has no kernel and so the Hilbert space $\hil$
admits an orthogonal decomposition
$$
\Hil=\Hil_+\oplus\Hil_-
$$
into the positive/negative eigenspaces $\Hil_\pm$ of the Dirac
operator $\Dirac_E$.

We can represent the commutative $C^*$-algebra $C(W)$ of worldvolume
functions on $\Hil$ by multiplication operators $M_f$ for $f\in
C(W)$. Generically the operator $M_f$ does not preserve the subspace
$\Hil_+$, but as before its compression to a Toeplitz operator
$\T_f=P_+\circ M_f:\Hil_+\to\Hil_+$ does, where now $P_+$ is the
orthogonal projection $\Hil\to\Hil_+$. As before, for $f,g\in C(W)$
the difference $\T_f\,\T_g-\T_{f\,g}$ is a compact operator on
$\Hil_+$ and so we get a Toeplitz extension
$$
0~\longrightarrow~\compact~\longrightarrow~\alg_W~
\longrightarrow~C(W)~\longrightarrow~0
$$
of the algebra of worldvolume functions, where $\alg_W$ is the
$C^*$-algebra generated by $\T_f$ for $f\in C(W)$. Let
$\tau_W:C(W)\to\Calkin(\Hil)$ be the corresponding Busby
invariant. Then by using the pullback $\zeta^*:C(X)\to C(W)$ we can
define a new Busby invariant
$\widetilde{\tau}:=\tau_W\circ\zeta^*:C(X)\to\Calkin(\Hil)$ which
corresponds to an extension
$$
0~\longrightarrow~\compact~\longrightarrow~\widetilde{\alg}~
\longrightarrow~C(X)~\longrightarrow~0
$$
of the algebra of functions on all of $X$. The corresponding
K-homology class in $\K_1^{\rm a}(X)$ is independent of the metric on
$X$ and of the choice of connection on $E$.
\end{proof}
\begin{remark}
The proof of the converse of the result proven here is beyond the
scope of these notes. The one-to-one correspondence between extension
classes in $\K_1^{\rm a}(X)$ and D-branes requires taking the quotient of the
set
of all Baum-Douglas K-cycles $(W,E,\zeta)$ by a collection of
equivalence relations~\cite{BD1,RSz1}. These relations make good
physical sense and the corresponding equivalence classes $[W,E,\zeta]$
capture the novel dynamical processes of D-brane
physics~\cite{HM1,AST1,RSz1,Sz1,LPSz1}. Note that in this
correspondence the D-brane worldvolume $W$ need not be a submanifold
of $X$. The problem of finding spaces $X$ for which the generators of
K-homology can be constructed from cycles of $X$ is related to the
Hodge conjecture. A projective algebraic variety that satisfies both
the requirements that cycles generate K-homology~\cite{RSz1} and the
hypothesis of the Hodge conjecture has certain restrictions on its
cohomology. A Calabi-Yau threefold satisfies these conditions,
and thus the K-cycles in physically viable string compactifications
$X$ correspond to D-branes whose worldvolumes are cycles of $X$.
\label{cyclerem}\end{remark}

Theorem~\ref{Dbranethm} provides us with the proper perspective on the
noncommutative solitons that we have constructed in this section. The
solitons provide the K-homology version of the equivalence of D-brane
charges in commutative and noncommutative field theories. This is
asserted via the equivalence between the analytic index and the
topological index via the Baum-Douglas K-cycle
construction~\cite{BD1,RSz1}. The Toeplitz operators on Hardy space
${\rm H}^2(\S^{2n-1},\dd\Omega)$ determine an algebra $\alg$ of
endomorphisms providing a non-trivial extension (\ref{extexactseq}) by
compact operators. This defines an analytic K-homology class in
$\K_1^{\rm a}(\S^{2n-1})$ which is the same as the class $[\Dirac]$
determined by the Dirac operator on $\S^{2n-1}$~\cite{BD1}. In
particular, Theorem~\ref{topchargethm} now follows
from the ordinary Atiyah-Singer index theorem. We may associate this
class with one in degree zero K-homology which is the appropriate
receptacle for the classification of D-brane charges in Type~IIB string
theory~\cite{WittenK,OSz1}. If we work in {\it relative}
K-homology~\cite{RSz1}, then the connecting
homomorphism in the six-term exact sequence for the pair $({\rm
  B}^{2n},\S^{2n-1}=\partial{\rm B}^{2n})$ yields an isomorphism
\beq
\partial\,:\,\K_0^{\rm a}\big({\rm B}^{2n}\,,\,\S^{2n-1}\big)~
\stackrel{\approx}{\longrightarrow}~\K_1^{\rm a}\big(\S^{2n-1}\big) \ .
\label{sixtermiso}\eeq
This determines an element of the compactly supported degree zero
K-homology of $\plane_{2n}$ associated to the noncommutative
soliton. We have in this way provided a description of D-branes in terms of
algebras $\alg$ of ``almost commuting'' operators corresponding to
``almost commutative'' spaces which extend the hyperplane
worldvolume $\plane_{2n}$.

\newsection{Gauge theory on ${\plane^\theta_{2n}}$\label{Gauge}}

In the previous section we have arrived at a description of
noncommutative solitons as D-branes in terms of K-cycles
$(W,E,\zeta)$. In the correspondence it was essential to introduce a
connection on the complex Chan-Paton vector bundle $E\to W$. It is
natural to now construct worldvolume field theories using these
connections. In the sequel we shall therefore focus our attention on
{\it gauge theories} on noncommutative spaces and examine what their
classical solutions can teach us. Worldvolume gauge theories are the
essence of the novel dynamical properties of D-branes in string
theory.

\subsection{Projective modules\label{projmodV}}

To construct gauge theory on $\plane_{2n}^\theta$ we proceed in the
usual way by introducing connections on projective modules over the
noncommutative algebra. The natural class of projective modules are
the collections of Fock modules
$\Fock^q:=\Fock\oplus\cdots\oplus\Fock$ ($q$ times). There are also
the trivial free modules of rank $N$ given by $N$ copies $\Hil^N$ of
the algebra $\Hil:=\plane_{2n}^\theta$ itself. Let us now go through some
general facts concerning projective modules over
$\plane_{2n}^\theta$~\cite{KSrev,Schwarz1}.
\begin{proposition}
Any finitely generated left projective module over the Moyal space
$\plane_{2n}^\theta$ is of the form $\mod_{N,q}=\Hil^N\oplus\Fock^q$
for some $N,q\in\nat_0$.
\label{modVprop}\end{proposition}
\begin{remark}
The integer $N$ corresponds to the rank, or zeroth Chern number
$c_0(\mod_{N,q})$, of the module. The integer $q$ corresponds to a
{\it topological charge} whose interpretation depends on dimension. For
example, when $n=1$ it is the {\it magnetic charge} or first Chern number
$c_1(\mod_{N,q})$, while when $n=2$ it is the {\it instanton number} or
second Chern number $c_2(\mod_{N,q})$.
\label{topchargerem}\end{remark}
\begin{cor}
The K-theory of the Moyal $2n$-space is given by
$\K_0(\plane_{2n}^\theta)=\zed\oplus\zed$ with positive cone
$\K_0^+(\plane_{2n}^\theta)=\nat\oplus\nat$.
\label{KtheoryVcor}\end{cor}
\begin{remark}
Any two projective modules representing the same element of K-theory
are isomorphic. The K-theory of $\plane_{2n}^\theta$ is quite different from
the
(compactly supported) K-theory of the ordinary topologically trivial
hyperplane $\plane_{2n}$. It allows for non-trivial topological
charges $q$ and so resembles more closely the K-theory of the sphere
$\S^{2n}$. This feature will be responsible later on for the
appearence of topologically non-trivial gauge field configurations
which have no counterparts in ordinary gauge theory on
$\plane_{2n}$. However, the positive cone $\K_0^+(\plane_{2n}^\theta)$
is different from that of $\S^{2n}$ as one cannot have stable modules
with negative charge $q<0$ in the present case. This is due to a
labelling problem, because there is no way to distinguish between the algebras
corresponding to the noncommutativity parameters $\theta$ and
$-\theta$ (see Remark~\ref{Equivrem}). This property will manifest
itself explicitly in the classical solutions that we shall construct
and its origin will be elucidated in Section~\ref{Decomp}. Physically, it will
imply that there is no way to produce vortices from anti-vortices on
$\plane_{2n}^\theta$ by simply changing the orientation of the
hyperplane.
\label{KtheoryVrem}\end{remark}

\subsection{Yang-Mills theory\label{YangMillsV}}

It is possible to define connections within the present class of
noncommutative spaces in the usual spirit and formalism
of noncommutative geometry~\cite{ConnesBook}. However, the noncommutative space
$\plane_{2n}^\theta$ has enough symmetries so that a simple definition
will suffice for our purposes. By a {\it connection} $\nabla$ on a
finitely-generated left projective $\plane_{2n}^\theta$-module $\mod$
we will mean a collection of anti-Hermitean $\complex$-linear operators
$\nabla_i:\mod\to\mod$, $i=1,\dots,2n$ satisfying the Leibniz rule
\beq
\nabla_i(f\cdot v)=\partial_i(f)\cdot v+f\cdot\nabla_i(v)
\label{LeibnizV}\eeq
for all $i=1,\dots,2n$, $v\in\mod$ and $f\in\plane_{2n}^\theta$. The
space of connections on $\mod$ is denoted ${\rm Conn}(\mod)$.

Let us find the general form of a connection on a generic projective
module as specified by Proposition~\ref{modVprop}. It is
straightforward to show that only {\it trivial} gauge fields arise on the
Fock module $\Fock$~\cite{GrossNek2}, in accordance with the
interpretation that $\Fock$ is like a single ``point'' on the
noncommutative space (see Remark~\ref{Fockpointrem}).
\begin{proposition}
If $\nabla$ is a connection on the Fock module $\Fock$, then
$$
\nabla_i=\Pi_m\circ\partial_i\circ\Pi_m+\alpha_i\,
\id \ , \quad i=1,\dots,2n
$$
for some $\alpha_i\in\complex$ and fixed
$m\in\nat_0$.
\label{trivialconnprop}\end{proposition}
\begin{proof}
{}From the Leibniz rule (\ref{LeibnizV}) and the inner derivation
property (\ref{partialinner}) one has the identity
$$
\nabla_i(f\cdot v)-f\cdot\nabla_i(v)=\ii\,\sum_{j=1}^{2n}\,
\bigl(\theta^{-1}\bigr)_{ij}\,\bigl[x^j\,,\,f\bigr]\cdot v
$$
which may be rewritten as
$$
\bigl[\nabla_i-\ii\,\mbox{$\sum\limits_{j=1}^{2n}$}\,(\theta^{-1})_{ij}\,
x^j\,,\,f\bigr]\cdot v=0
$$
for all $f\in\plane_{2n}^\theta$ and all $v\in\Fock$. It follows that
the operator $\nabla_i-\ii\sum_j\,(\theta^{-1})_{ij}\,x^j$ lives in
the center of the algebra ${\rm
  End}_{\plane_{2n}^\theta}(\Fock)\cong\plane_{2n}^\theta$ and hence
is proportional to the identity endomorphism $\one$ of~$\Fock$.
\end{proof}

We can construct {\it non-trivial} gauge fields instead on the trivial
module given by the algebra $\Hil=\plane_{2n}^\theta$ itself. By the
Leibniz rule (\ref{LeibnizV}) any connection $\nabla_i:\Hil\to\Hil$
can be written in the form
\beq
\nabla_i(v)=-\ii\sum_{j=1}^{2n}\,\bigl(\theta^{-1}\bigr)_{ij}\,x^j\cdot
v +D_i(v)
\label{connHilgen}\eeq
for $v\in\Hil$, where $D_i\in{\rm End}(\Hil)$ is any anti-Hermitean
operator on $\Hil$. We choose
\beq
D_i=\ii\sum_{j=1}^{2n}\,\bigl(\theta^{-1}\bigr)_{ij}\,x^j+A_i
\label{Dichoice}\eeq
where $A_i\in{\rm End}_{\plane_{2n}^\theta}(\Hil)$ are
anti-Hermitean $\plane_{2n}^\theta$-linear endomorphisms. As all Moyal
spaces $\plane_{2n}^\theta$, $\theta\in\real\,\setminus\,\{0\}$ are Morita
equivalent (in fact isomorphic), $A_i$ can be taken to be
anti-Hermitean elements of the algebra $\plane_{2n}^\theta$
itself. Then by the inner derivation property (\ref{partialinner}) one
has
\beq
\nabla_i(v)=\partial_i(v)+A_i\cdot v \ .
\label{connHilchoice}\eeq

As usual, we define the {\it curvature} of a connection to be a
measure of the deviation of the mapping $\partial_i\mapsto\nabla_i$ from
being a homomorphism of the Lie algebra (\ref{translgprep}) of
automorphisms of $\plane_{2n}^\theta$. This gives the collection of
anti-Hermitean endomorphisms $F_{ij}\in{\rm
  End}_{\plane_{2n}^\theta}(\Hil)\cong\plane_{2n}^\theta$,
$i,j=1,\dots,2n$ defined by
\beq
F_{ij}=\big[\nabla_i\,,\,\nabla_j\big]=\big[D_i\,,\,D_j\big]
-\ii\big(\theta^{-1}\big)_{ij}\,\one \ .
\label{curvVdef}\eeq
The {\it Yang-Mills action functional} ${\rm YM}:{\rm
  Conn}(\Hil)\to[0,\infty)$ is defined by
\beq
{\rm YM}(\nabla)=-\mbox{$\frac14$}\,
\Tr\Bigl[\,\sum_{i,j=1}^{2n}\,\bigl(F_{ij}\bigr)^2\,\Bigr]
=-\mbox{$\frac14$}\,
\Tr\Bigl[\,\sum_{i,j=1}^{2n}\bigl([D_i,D_j]-\ii(\theta^{-1})_{ij}
\,\one\bigr)^2\,\Bigr] \ .
\label{YMVdef}\eeq
The Yang-Mills functional (\ref{YMVdef}) is invariant under the {\it
  gauge transformations}
\beq
A_i~\longmapsto~U\,\partial_i\bigl(U^{-1}\bigr)+U\,A_i\,U^{-1} \ ,
\quad i=1,\dots,2n
\label{AiVgaugetransf}\eeq
with $U\in{\rm U}(\Hil)$, which induce the unitary transformations
\beq
D_i~\longmapsto~U\,D_i\,U^{-1} \ , \quad
F_{ij}~\longmapsto~U\,F_{ij}\,U^{-1} \ .
\label{gaugetransfVrem}\eeq

\subsection{Fluxons\label{Fluxons}}

We can explicitly construct all exact solutions to Yang-Mills theory on the
Moyal plane $\plane_2^\theta$, and hence for the remainder of this
section we will focus on this case. It is convenient to introduce
formal complex combinations of the operators (\ref{Dichoice})
using the basic elements (\ref{creandef}) to write
\beq
D=-\mbox{$\frac1{\sqrt{2\,\theta}}$}\,a^\dag+\mbox{$\frac12$}\,
(A_1-\ii A_2) \ , \quad \overline{D}=
\mbox{$\frac1{\sqrt{2\,\theta}}$}\,a+\mbox{$\frac12$}\,
(A_1+\ii A_2) \ .
\label{DbarDdef}\eeq
In terms of these operators the curvature (\ref{curvVdef}) reads
\beq
F:=F_{12}=2\ii\left(\bigl[\,\overline{D}\,,\,D\bigr]-\mbox{$
\frac1{2\,\theta}$}\,\one\right)
\label{curvDbarD}\eeq
and the Yang-Mills functional (\ref{YMVdef}) becomes
\beq
{\rm YM}\bigl(\nabla\bigr)={\rm YM}\bigl(D\,,\,\overline{D}\,
\bigr):=2\,\Tr\left(\bigl[\,\overline{D}\,,\,D\bigr]-\mbox{$
\frac1{2\,\theta}$}\,\one\right)^2 \ .
\label{YMDbarD}\eeq

Applying the variational principle to the action (\ref{YMDbarD})
yields the {\it Yang-Mills equations} on $\plane_2^\theta$ given by
\beq
\bigl[D\,,\,\bigl[\,\overline{D}\,,\,D\bigr]\bigr]=
\bigl[\,\overline{D}\,,\,\bigl[\,\overline{D}\,,\,D\bigr]\bigr]=0  \ .
\label{YMeqsV}\eeq
As previously, we are interested in special classes of solutions to
these equations.
\begin{definition}
Let $q\in\nat$. A {\it $q$-fluxon} on $\plane_2^\theta$ is an
anti-Hermitean solution $D,\overline{D}\in{\rm End}(\Hil)$ of the
Yang-Mills equations (\ref{YMeqsV}) which has topological charge
$c_1(\Hil)=\Tr(F)=q$ and for which the Yang-Mills functional ${\rm
  YM}(D,\overline{D}\,)$ is well-defined and finite.
\label{fluxondef}\end{definition}
\noindent
A fluxon solution is a soliton on the noncommutative plane carrying a
magnetic charge or ``flux''~\cite{Poly1,GrossNek2}. Using the shift
operator $\shift$ introduced in Example~\ref{Toeplitzex}, it is
straightforward to explicitly construct all finite action solutions of
the Yang-Mills equations on the Moyal plane
$\plane_2^\theta$~\cite{GrossNek2}.
\begin{theorem}
To each collection
$\lambda_0,\lambda_1,\dots,\lambda_{q-1}\in\complex$ of fixed complex
numbers there bijectively corresponds a $q$-fluxon
$$
D^{(q)}_{\{\lambda_i\}}
=\sum_{i=0}^{q-1}\,\lambda_i~e_i\otimes e_i^*-\big(\shift
\big)^q\,c^\dag\,\big(\shift^\dag\big)^q \ , \quad
\overline{D}{\,}^{(q)}_{\{\lambda_i\}}
=\sum_{i=0}^{q-1}\,\overline{\lambda}_i~e_i\otimes e_i^*+
\big(\shift\big)^q\,c\,\big(\shift^\dag\big)^q
$$
in ${\rm End}(\Hil)$ of action
$$
{\rm YM}\bigl(D^{(q)}_{\{\lambda_i\}}\,,\,
\overline{D}{\,}^{(q)}_{\{\lambda_i\}}\bigr)=2\pi\,\theta^{-1}\,q \ ,
$$
where $c,c^\dag$ generate the irreducible representation of the
Heisenberg algebra.
\label{fluxonthm}\end{theorem}
\begin{proof}
We need to find a pair of anti-Hermitean operators $D,\overline{D}$
on the free module $\Hil$ over $\plane_2^\theta$ which obey the
equations
$$
\bigl[D\,,\,\overline{D}\,\bigr]=\mbox{$\frac12\,\bigl(\frac1\theta\,
\one+\ii F\bigr)$} \ , \quad \bigl[D\,,\,F\bigr]=\bigl[\,
\overline{D}\,,\,F\bigr]=0 \ .
$$
These equations imply that $D,\overline{D},\frac12\,\big(\frac1\theta+\ii
F\big)$ form a representation of the Heisenberg commutation relations,
with the curvature $F$ generating the center of the algebra. Under the
action of these operators, the module $\Hil$ thereby decomposes as
$$
\Hil=\bigoplus_{n}\,\Hil_n
$$
into irreducible representations $\Hil_n\cong\Fock$ of this Heisenberg
algebra. Then for each $n$ one has $F|_{\Hil_n}=f_n\,\one$ for some
$f_n\in\ii\real$. Set $d_n:=\dim(\Hil_n)$. Then $d_n$ is
infinite unless $1+\ii\theta\,f_n=0$. The finite action constraint
requires $\Tr(F^2)<\infty$, where
$$
\Tr\bigl(F^2\bigr)=\sum_{n}\,d_n\,f_n^2 \ .
$$
This is a sum of negative terms. If $d_n$ is infinite for some
$n$, then the only way to make this quantity well-defined is
to have $f_n=0$ in such a way that the regulated trace yields a finite
product $d_n\,f_n^2$. On the other hand, if some $d_n\in\nat$ then
$f_n=-\frac\ii\theta$, and the finite action condition implies that
there are only finitely many such positive dimensions.

These facts imply that the fluxon solution is determined by a
finite-dimensional linear subspace $V_q\subset\Hil$ which may be
characterized as follows. Via a gauge transformation if necessary, we
may assume that $V_q$ is the linear span of the number basis vectors
$e_0,e_1,\dots,e_{q-1}$. On $V_q$, the operators $D$ and
$\overline{D}$ commute, $[D,\overline{D}\,]=0$, and without loss of
generality may be taken to be diagonal operators with respect to
the chosen basis of $V_q$ so that
$$
D\bigm|_{V_q}=\sum_{i=0}^{q-1}\,\lambda_i~e_i\otimes e_i^* \ , \quad
\overline{D}\,\bigm|_{V_q}=\sum_{i=0}^{q-1}\,\overline{\lambda}_i
{}~e_i\otimes e_i^*
$$
for some fixed
$\lambda_0,\lambda_1,\dots,\lambda_{q-1}\in\complex$. On the
orthogonal complement $\Hil\ominus V_q\cong\Hil$, there exists
$N\in\nat$ such that the operators
are instead generically given by a reducible sum of $N$ irreducible
representations of the Heisenberg algebra as
$$
D\bigm|_{\Hil\ominus
  V_q}=\bigoplus_{k=0}^{N-1}\,\bigl(-c_{(k)}^\dag\bigr) \ , \quad
\overline{D}\,\bigm|_{\Hil\ominus
  V_q}=\bigoplus_{k=0}^{N-1}\,\bigl(c_{(k)}\bigr)
$$
with $[c_{(k)},c_{(k)}^\dag]=1$ for each $k=0,1,\dots,N-1$. By the
Stone-von~Neumann theorem, $c_{(k)}=c$ and $c_{(k)}^\dag=c^\dag$ for
each $k$ and hence
$$
D\bigm|_{\Hil\ominus V_q}=-c^\dag\otimes\one_N \ , \quad
\overline{D}\,\bigm|_{\Hil\ominus V_q}=
c\otimes\one_N \ .
$$
This makes the Hilbert space $\Hil\ominus V_q$ into an $N$-fold Fock
module $\Fock^N\cong\ell^2(\nat_0^N)$ with number basis $e_n^{(k)}$,
$n\in\nat_0$, $k=0,1,\dots,N-1$ defined by the actions
\beq
c^\dag\cdot e_n^{(k)}&=&\sqrt{n+1}~e_{n+1}^{(k)} \ , \nonumber\\
c\cdot e_n^{(k)}&=&\sqrt n~e_{n-1}^{(k)} \ , \nonumber\\
\one_N\cdot e_n^{(k)}&=&e_n^{(k)} \ . \nonumber
\eeq

Let us take $N=1$. Let $\shift_q{}^\dag:\Hil\ominus V_q\to\Hil$ be a
unitary isomorphism of $\plane_2^\theta$-modules. Extend
$\shift_q{}^\dag$ to all of $\Hil$ by setting it equal to~$0$ on
$V_q$. Then as an operator on $\Hil$ it satisfies
$$
\shift_q{}^\dag\,\shift_q=\one \ , \quad
\shift_q\,\shift_q{}^\dag=\one-\P_q
$$
where $\P_q$ is the orthogonal projection $\Hil\to V_q$. In other
words, the endomorphism $\shift_q\in{\rm End}(\Hil)$ is a partial
isometry of $\Hil$. With respect to the chosen number basis, we have
$\P_q=\P_{(q)}=\sum_{i=0}^{q-1}\,e_i\otimes e_i^*$ and
$\shift_q=(\shift)^q$ where $\shift$ is the shift endomorphism
introduced in Example~\ref{Toeplitzex}. The conclusion now follows
from computing the corresponding curvature to get
$F=\frac\ii\theta\,\P_q$.
\end{proof}
\begin{remark}
The $q$-fluxon is labelled by the set of moduli
$\lambda_i\in\complex$, $i=0,1,\dots,q-1$ which describe the
positions or separations of the {\it vortices} (carrying magnetic
charge $q\in\nat$) on $\plane_2$~\cite{GrossNek2}. The explicit
solution constructed
above is of rank~$1$. Higher rank fluxons can be similarly constructed
by choosing $N>1$ in the proof of Theorem~\ref{fluxonthm}, with no
qualitative change by the Hilbert hotel argument. Thus, in
addition to their moduli, fluxons are labelled by K-theory charges
$(N,q)\in\K_0^+(\plane_2^\theta)$. Identifying the gauge equivalence
classes of fluxon solutions now consists in quotienting by the
discrete Weyl subgroup $S_q\subset{\rm U}(q)\subset{\rm U}(\Hil)$
acting non-trivially on the subspace $V_q$ above by permuting the
fluxon positions $\lambda_i$.
\label{fluxonmodrem}\end{remark}
\begin{cor}
The moduli space $\mathcal{G}_{N,q}(\plane_2^\theta)$ of fluxons of
K-theory charge $(N,q)$ is the $q$-th symmetric product orbifold
$$
\mathcal{G}_{N,q}\big(\plane_2^\theta\big)~=~{\rm Sym}^q\big(\plane_2\big)
\ .
$$
\label{fluxonmodcor}\end{cor}
\begin{remark}
The value of the Yang-Mills functional on a fluxon as in
Theorem~\ref{fluxonthm} diverges in the formal limit $\theta\to0$,
consistent with the lack of finite action topologically non-trivial
field configurations in ordinary gauge theory on the plane
$\plane_{2}$. Note that the fluxon solution appears with only one
sign of the topological charge $q$, consistent with the K-theory description of
Section~\ref{projmodV}. These configurations are not global minima of
the Yang-Mills functional in general and hence are {\it unstable}. In
string theory they can be interpreted as describing $q$ unstable
D0-branes inside $N$ D2-branes with worldvolume $\plane_2$, in a
background $B$-field and in the Seiberg-Witten
limit~\cite{GrossNek1,GrossNek2,AGMS1}.
\label{fluxonstringrem}\end{remark}

\newsection{Toroidal D-branes\label{TorD}}

In the sequel we will leave the setting of Moyal spaces and start
looking at more complicated worldvolume geometries. As the simplest
extension of our previous considerations, in this section we
will still work with flat target spaces $X$ but we will assume that
the worldvolume is compactified on a two-dimensional torus
$\Torus^2$. This has the effect of bringing in non-trivial topological
effects while still retaining the relative simplicity of flat
worldvolume geometries. The presence of the constant $B$-field deforms
the worldvolume to a noncommutative torus $\Torus_\Theta^2$ which is
the most studied and best understood example of a noncommutative
space~\cite{ConnesBook}. After a quick reminder of some pertinent
aspects of the noncommutative geometry of this space, we will
construct all (finite action) solutions of the corresponding
Yang-Mills equations. We will then show that these solutions possess a
remarkably intimate connection with the fluxon solutions constructed
in the previous section.

\subsection{Solitons on the noncommutative torus\label{NCTorus}}

The two-dimensional noncommutative torus $\Torus_\Theta^2$ is the
classic example of a noncommutative space and we will only briefly mention
some facets of its geometry, primarily to set notation. Unless
otherwise explicitly stated, we will fix an irrational number
$\Theta\in(0,1)\,\cap\,(\real\,\setminus\,\rat)$. We define
$\Torus_\Theta^2$ to be the associative unital $*$-algebra generated
by a pair of unitaries $U_1,U_2$ obeying the single relation
\beq
U_1\,U_2=\e^{2\pi\ii\Theta}~U_2\,U_1 \ .
\label{NCtorusdef}\eeq
As with Moyal spaces, we will regard $\Torus^2_\Theta$ as an
appropriate ``algebra of Schwartz functions'' defined in terms of
expansions in the generators $U_1,U_2$. There is again a natural trace
$\Tr:\Torus_\Theta^2\to\complex$ on the algebra which is positive and
faithful, and hence a natural notion of integral. A collection of
linear derivations $\partial_i:\Torus^2_\Theta\to\Torus^2_\Theta$ may
be defined on generators by
\beq
\partial_i(U_j)=2\pi\ii\delta_{ij}~U_i \ , \quad i,j=1,2
\label{derivT2}\eeq
and extended as automorphisms of $\Torus^2_\Theta$ by linearity and
the Leibniz rule. As in the Moyal case these derivations commute,
\beq
\bigl[\partial_i\,,\,\partial_j\bigr]=0 \ ,
\label{derivT2comm}\eeq
and they generate the action of the two-dimensional translation group of
$\Torus^2$ on the noncommutative algebra. There is also a Weyl-Wigner
correspondence completely analogous to that of Section~\ref{Defquant}
which may be used to define $\Torus_\Theta^2$ via deformation
quantization~\cite{Szrev}, but we will not need this formalism here.

It is possible to now proceed in an analogous way to
Section~\ref{Solitons} with the construction of scalar field theories
on $\Torus_\Theta^2$ and their soliton
solutions~\cite{MartMoore1,BKMT1,KMT1,KrajSch1}. The resulting
configurations are similar to those of Moyal spaces, except that the
non-trivial topology now leads to far richer structures. Explicit
projector solitons are provided by the Powers-Rieffel projector on
$\Torus_\Theta^2$~\cite{Rieffel1}. Its Wigner function is not a
localized solitonic ``lump'' like the Gaussian soliton field configurations on
$\plane_{2n}^\theta$. Rather, it is localized along one direction but
yields stripe-like patterns in the other direction of
$\Torus^2$~\cite{BKMT1,LLSz2}. On the other hand, the homotopically
equivalent Boca projector~\cite{Boca1} more
closely resembles the Gaussian projector solitons on
$\plane_{2n}^\theta$~\cite{KrajSch1,LLSz2}. Partial isometry solitons
on $\Torus^2_\Theta$ can also be constructed as the ``angular''
operators appearing in the polar decompositions of closed bounded
operators based on the Powers-Rieffel projection~\cite{LLSz2}, in
analogy to the angular dependence of the
Wigner functions for the basic shift partial isometry $\shift$ on
$\plane_{2n}^\theta$ (see~Example~\ref{starprodex}). The explicit
descriptions of these configurations are much more involved than in the
Moyal case and are beyond the scope of these notes. Instead, we will
now proceed to focus our attention to gauge theories on the
noncommutative torus.

\subsection{Gauge theory\label{YMTorus}}

We begin by describing the K-theory of the noncommutative
torus~\cite{ConnesBook}.
\begin{proposition}
The K-theory of the noncommutative torus is given by the ordered subgroup
$\K_0(\Torus_\Theta^2)=\zed+\zed\,\Theta$ of $\real$ with positive cone
$\K_0^+(\Torus_\Theta^2)=\{(p,q)\in\zed^2~|~p-q\,\Theta>0\}$.
\label{K0T2prop}\end{proposition}
\begin{proof}
K-theory is stable under deformations of unital algebras so
$\K_0(\Torus_\Theta^2)\cong\K_0(C(\Torus^2))=\zed\oplus\zed$. For
$(p,q)\in\zed^2$, let $\P=\P_{p,q}\in\mat_n(\Torus_\Theta^2)$,
$n\in\nat$ be a projector representing the corresponding
Murray-von~Neumann equivalence class. One has
$$
\Tr\bigl(\P_{p,q}\bigr)=\Tr\bigl(\P_{p,q}\,\P_{p,q}{}^\dag
\bigr)=p-q\,\Theta
$$
and hence the trace yields an isomorphism
$\Tr:\K_0(\Torus_\Theta^2)\stackrel{\approx}{\longrightarrow}
\zed+\zed\,\Theta$, with the trivial identity projector $\one$
generating the first copy of $\zed$ and the Powers-Rieffel projector
generating the second copy of $\zed$. The trace of $\P_{p,q}$
coincides with the Murray-von~Neumann dimension of its image
subspace.
\end{proof}
\noindent
For $(p,q)\in\zed^2$, the space
$\mod_{p,q}=\P_{p,q}(\Torus_\Theta^2)^n$ is a finitely generated left
projective module over $\Torus_\Theta^2$ called a {\it Heisenberg
  module}.
\begin{proposition}
Any finitely generated projective module over the
noncommutative torus $\Torus_\Theta^2$ which is not free is isomorphic
to a Heisenberg module $\mod_{p,q}$.
\label{Heisenmodprop}\end{proposition}

Henceforth we will consider only Heisenberg modules over the
noncommutative torus. The module $\mod_{p,q}$ is a stable module
labelled by the positive cone $(p,q)\in\K_0^+(\Torus_\Theta^2)$ of the
K-theory group and has positive Murray-von~Neumann dimension
\beq
\dim(\mod_{p,q})=\Tr(\P_{p,q})=p-q\,\Theta>0 \ .
\label{dimmodpq}\eeq
The integer
\beq
q=\mbox{$\frac1{2\pi\ii}$}\,\Tr\left[\P_{p,q}\,\bigl(\partial_1(
\P_{p,q})\,\partial_2(\P_{p,q})-\partial_2(
\P_{p,q})\,\partial_1(\P_{p,q})\bigr)\right]
\label{qChern}\eeq
is the Chern number $c_1(\mod_{p,q})$ and will be referred to as the
{\it topological charge} of the Heisenberg module. We define the {\it
  rank} $N$ of $\mod_{p,q}$ to be the positive integer
\beq
N={\rm gcd}(p,q) \ .
\label{rankmodpq}\eeq
We will always assume $q>0$ in what follows.

We define connections $\nabla$ on Heisenberg modules analogously to
the Moyal case, i.e. as pairs of anti-Hermitean $\complex$-linear
operators $\nabla_1,\nabla_2:\mod_{p,q}\to\mod_{p,q}$ obeying the
usual Leibniz rule analogously to eq.~(\ref{LeibnizV}). The space of
connections on a Heisenberg module is denoted ${\rm
  Conn}(\mod_{p,q})$. We can use the automorphisms $\partial_i$ given
by eq.~(\ref{derivT2}) to define fixed fiducial connections and write
$\nabla_i=\P_{p,q}\circ\partial_i\circ\P_{p,q}+A_i$ with
$A_i\in{\rm End}_{\Torus_\Theta^2}(\mod_{p,q})$. The curvature
$F\in{\rm End}_{\Torus_\Theta^2}(\mod_{p,q})$ of a
connection $\nabla$ is defined as usual by
\beq
F:=\bigl[\nabla_1\,,\,\nabla_2\bigr] \ ,
\label{curvdefT2}\eeq
and one may introduce the Yang-Mills functional ${\rm YM}:{\rm
  Conn}(\mod_{p,q})\to[0,\infty)$ on a {\it fixed} Heisenberg module
as
\beq
{\rm YM}(\nabla)=-\mbox{$\frac\tau2$}\,\Tr\bigl([\nabla_1,\nabla_2]^2
\bigr)=-\mbox{$\frac\tau2$}\,\Tr\bigl(F^2\bigr) \ .
\label{YMdefT2}\eeq
For later convenience we have introduced a constant parameter
$\tau>0$ which we identify as the (imaginary part of the) modulus of
the torus $\Torus^2$. It should accompany all trace operations in
order to keep quantities consistently defined. The Yang-Mills
functional (\ref{YMdefT2}) is invariant under the gauge
transformations
\beq
\nabla_i~\longmapsto~U\,\nabla_i\,U^{-1} \ , \quad i=1,2
\label{nablagaugetransf}\eeq
with $U\in{\rm U}(\mod_{p,q})$.
\begin{proposition}
Any Heisenberg module $\mod_{p,q}$ admits a connection $\nabla^{\rm
  c}$ of constant curvature
$$
F_{p,q}:=\bigl[\nabla_1^{\rm c}\,,\,\nabla_2^{\rm c}\bigr]=
\frac{2\pi\ii}\tau\,\frac q{p-q\,\Theta}~\P_{p,q} \ .
$$
\label{constcurvprop}\end{proposition}
\begin{remark}
Since $\mod_{p,q}=\P_{p,q}(\Torus_\Theta^2)^n$ for some $n\in\nat$ and
$(\P_{p,q})^2=\P_{p,q}$, the projector $\P_{p,q}$ acts as the identity
endomorphism on the Heisenberg module and hence the curvature
$F_{p,q}$ is ``constant''. It now follows that the topological charge
(\ref{qChern}) can be expressed in the standard form
$$
q=\mbox{$\frac\tau{2\pi\ii}$}\,\Tr(F_{p,q}) \ .
$$
The equality of the two expressions for the topological charge is
essentially the index theorem.
\label{constcurvrem}\end{remark}

Using Proposition~\ref{constcurvprop} we can obtain an explicit
description of the Heisenberg module $\mod_{p,q}$ as the separable
Hilbert space
\beq
\mod_{p,q}=\Fock\otimes\Weyl_{p,q} \ .
\label{modpqexpl}\eeq
The Fock module $\Fock$ is the irreducible representation of the
Heisenberg commutation relation $[\nabla_1^{\rm c},\nabla_2^{\rm
  c}]=F_{p,q}$, while $\mathcal{W}_{p,q}\cong\complex^q$ is the
$q\times q$ representation of the Weyl algebra
\beq
\Gamma_1\,\Gamma_2=\e^{2\pi\ii p/q}~\Gamma_2\,\Gamma_1
\label{Weylalgdef}\eeq
whose unique irreducible module has rank $\frac{q}{{\rm
    gcd}(p,q)}=\frac qN$. The generators
$U_1,U_2$ of the noncommutative torus then act on $\mod_{p,q}$ as the
operators
\beq
U_i=\exp\bigl(\mbox{$\frac{\sqrt\tau}q$}\,(p-q\,\Theta)\,
\nabla_i^{\rm c}\bigr)\otimes\Gamma_i \ , \quad i=1,2 \ .
\label{Uimodpq}\eeq

\subsection{Instantons\label{Instantons}}

Varying the Yang-Mills functional (\ref{YMdefT2}) in the usual way
leads to the Yang-Mills equations
\beq
\bigl[\nabla_1\,,\,\bigl[\nabla_1\,,\,\nabla_2\bigr]\bigr]=
\bigl[\nabla_2\,,\,\bigl[\nabla_1\,,\,\nabla_2\bigr]\bigr]=0
\label{YMeqsT2}\eeq
on the Heisenberg module $\mod_{p,q}$. As always, we are interested in
the finite action solutions to these equations up to gauge equivalence
defined by the transformations (\ref{nablagaugetransf}).
\begin{definition}
An {\it instanton} of K-theory charge
  $(p,q)\in\K_0^+(\Torus_\Theta^2)$ on $\Torus_\Theta^2$ is a
solution $\nabla\in{\rm
  Conn}(\mod_{p,q})$ of the Yang-Mills equations (\ref{YMeqsT2}) on
the Heisenberg module $\mod_{p,q}$ for which the Yang-Mills functional
${\rm YM}(\nabla)$ is well-defined and finite.
\label{instantonT2def}\end{definition}
\noindent
The explicit construction of such solutions is more involved than on
the Moyal plane $\plane_2^\theta$, because in the present case the
derivations (\ref{derivT2}) are {\it outer} automorphisms of the
noncommutative algebra.

An obvious solution to the Yang-Mills equations (\ref{YMeqsT2}) is
the constant curvature connection $\nabla=\nabla^{\rm c}$ on
$\mod_{p,q}$ of topological charge $q$. It has action
\beq
\YM(\nabla^{\rm c})=\frac{4\pi^2}\tau\,\frac{q^2}{p-q\,\Theta} \ ,
\label{YMconstcurv}\eeq
and one can show that this is the absolute minimum value of the
Yang-Mills functional on ${\rm
  Conn}(\mod_{p,q})$~\cite{ConnesRieffel1}.
\begin{proposition}
$\displaystyle \ \YM(\nabla^{\rm c})=\inf_{\nabla\in{\rm
    Conn}(\mod_{p,q})}\,\YM(\nabla) \ . $
\label{globalprop}\end{proposition}
\begin{remark}
Proposition~\ref{globalprop} implies that the constant curvature
connections define {\it stable} vacuum states in noncommutative gauge
theory. In the string theory setting, they correspond to $\frac12$-BPS
configurations~\cite{KSrev}.
\label{BPSrem}\end{remark}

The remaining instanton solutions to eqs.~(\ref{YMeqsT2}) are {\it
  unstable} and may be completely classified as
follows~\cite{Rieffel2,PanSz1}.
\begin{definition}
A {\it partition} of the K-theory charge
$(p,q)\in\K_0^+(\Torus_\Theta^2)$ is a collection
$\underline{(p,q)}=\{(p_k,q_k)\}$ of charges
$(p_k,q_k)\in\K_0^+(\Torus_\Theta^2)$ for which
$$
(p,q)=\sum_k\,(p_k,q_k) \ .
$$
\label{partitiondef}\end{definition}
\begin{theorem}
To each partition
$\underline{(p,q)}=\{(p_k,q_k)\}$ of $(p,q)\in\K_0^+(\Torus_\Theta^2)$
with finitely many components $(p_k,q_k)$ there bijectively
corresponds an instanton $\nabla=\nabla_{\underline{(p,q)}}$ of
K-theory charge $(p,q)$ with action
$$
\YM\bigl(\,\nabla_{\underline{(p,q)}}\,\bigr)=\frac{4\pi^2}\tau\,\sum_k\,
\frac{q_k^2}{p_k-q_k\,\Theta} \ .
$$
\label{instsolsthm}\end{theorem}
\begin{proof}
The idea of the proof is similar to that of
Theorem~\ref{fluxonthm}. An instanton $\nabla$ satisfies the equations
$$
\bigl[\nabla_1\,,\,\nabla_2\bigr]=F \ , \quad
\bigl[\nabla_1\,,\,F\bigr]=\bigl[\nabla_2\,,\,F\bigr]=0
$$
which describes a Heisenberg algebra with generators $\nabla_1$,
$\nabla_2$ and $F$ giving the central element. Under the action of
these operators, the Heisenberg module $\mod_{p,q}$ thus decomposes
into irreducible representations of this algebra to give
$$
\mod_{p,q}=\bigoplus_k\,\mod_{p_k,q_k}
$$
for some $(p_k,q_k)\in\zed^2$, with each submodule
$\mod_{p_k,q_k}$ a Heisenberg module over
$\Torus_\Theta^2$, i.e. $p_k-q_k\,\Theta>0$. By the Yang-Mills
equations, a solution $\nabla:\mod_{p_k,q_k}\to\mod_{p_k,q_k}$
preserves the subspaces $\mod_{p_k,q_k}$. As a consequence,
$\nabla_{(k)}^{\rm c}:=\nabla|_{\mod_{p_k,q_k}}$ has constant
curvature $F_{p_k,q_k}=F|_{\mod_{p_k,q_k}}$. The Yang-Mills functional
is additive,
$$
\YM\bigl(\,\mbox{$\bigoplus\limits_k\,\nabla_{(k)}$}\,\bigr)
=\sum_k\,\YM\bigl(\nabla_{(k)}\bigr) \ ,
$$
and by Proposition \ref{globalprop} the constant curvature connection
$\nabla_{(k)}=\nabla_{(k)}^{\rm c}$ is the global minimum of
$\YM\big|_{\mod_{p_k,q_k}}$ for each $k$.

Thus a critical point of the Yang-Mills functional on this direct sum
decomposition is given by
$$
\nabla=\nabla_{\underline{(p,q)}}:=\bigoplus_k\,\nabla_{(k)}^{\rm c} \ ,
$$
and every classical solution of noncommutative Yang-Mills theory is
characterized by module splittings in this way. The value of
$\YM(\,\nabla_{\underline{(p,q)}}\,)=\sum_k\,\YM(\nabla_{(k)}^{\rm
  c})$ is given from eq.~(\ref{YMconstcurv}). Since each term in this
sum is positive, the sum is finite only if there are finitely many
terms in the decomposition. Since the module decomposition is an
orthogonal direct sum, one has
$\dim(\mod_{p,q})=\sum_k\,\dim(\mod_{p_k,q_k})$ which is equivalent to
$$
p-q\,\Theta=\sum_k\,(p_k-q_k\,\Theta) \ .
$$
Since $\Theta$ is an irrational number, this is equivalent to the
partition conditions.
\end{proof}
\begin{remark}
The conservation of topological charge
$$
q=\sum_k\,q_k
$$
arising from the partition condition characterizes Yang-Mills theory
on a fixed Heisenberg module $\mod_{p,q}$. If we extrapolate these
solutions to the commutative case $\Theta=0$ of ordinary gauge theory
on $\Torus^2$, then this theory is distinguished from ``physical''
Yang-Mills theory which would sum over all topological
charges~\cite{PanSz1,PanSz2}. In the
noncommutative case only the fixed module definition of Yang-Mills
theory is well-defined, as there is no way in this case to naturally
``separate'' the topological numbers $p$ and $q$. Note that the global
minimum of the Yang-Mills
functional corresponds to the special case of the trivial partition
$\underline{(p,q)}=\{(p,q)\}$ having only a single component.
\label{qpartrem}\end{remark}
\begin{remark}
Since $\underline{(p,q)}$ contains only finitely many components
(although it can have an arbitrarily large number), one
can pick out from the corresponding module decomposition the submodule
of minimum dimension and thus order the partition components according to
increasing Murray-von~Neumann dimension~\cite{PanSz1}. The set of values of the
Yang-Mills action on its critical point set (the set of all partitions
or instantons) is discrete, and thus $\YM$ defines a Morse functional
on the space of connections $\Conn(\mod_{p,q})$~\cite{Rieffel2}.
\end{remark}

\subsection{Instanton moduli spaces\label{Instmodspace}}

It is straightforward to construct the moduli spaces of the solutions
obtained in Section~\ref{Instantons}. We first consider the moduli
space $\mathcal{G}_{(p,q)}(\Torus_\Theta^2)$ of constant curvature
connections $\nabla^{\rm c}$ on $\mod_{p,q}$
corresponding to the global minima of the Yang-Mills
functional~\cite{ConnesRieffel1,PanSz1}.
\begin{theorem}
The moduli space $\mathcal{G}_{(p,q)}(\Torus_\Theta^2)$ of stable
instantons on the Heisenberg module $\mod_{p,q}$ is the $N$-th
symmetric product orbifold
$$
\mathcal{G}_{(p,q)}\bigl(\Torus_\Theta^2\bigr)~=~{\rm Sym}^N\bigl(
\tilde\Torus{}^2\bigr)
$$
of the dual torus $\tilde\Torus{}^2$, where $N={\rm gcd}(p,q)$ is the
rank of $\mod_{p,q}$.
\label{globalmodspthm}\end{theorem}
\begin{proof}
Write the realization (\ref{modpqexpl}) of the Heisenberg module as
$$
\mod_{p,q}=\Fock\otimes\bigl(\Weyl_{\zeta_1}\oplus\cdots
\oplus\Weyl_{\zeta_N}\bigr)
$$
where, for each $i=1,\dots,N$, the module
$\Weyl_{\zeta_i}\cong\complex^{q/N}$ is the irreducible representation
of the Weyl algebra (\ref{Weylalgdef}) with central generator
$\zeta_i\in\tilde\Torus{}^2$. The only gauge transformations
(\ref{nablagaugetransf}) which act non-trivially on this decomposition
live in the Weyl subgroup $S_N\subset{\rm U}(N)\subset{\rm
  U}(\mod_{p,q})$ and act by permuting the various components
$\Weyl_{\zeta_i}$ of the direct sum. Dividing by this subgroup gives
the desired moduli space.
\end{proof}
\begin{remark}
As in the case of fluxons, the stable instanton moduli space coincides
with the quantum mechanical configuration space of a number of identical
particles on $\tilde\Torus{}^2$. In this sense the gauge theory is
``topological''~\cite{PanSz1,PanSz3}, in that it resembles more
closely quantum mechanics rather than field theory. Notice that the
number of instantons here is determined by the rank $N$ of the gauge
theory, unlike the fluxon number which is determined by the
topological charge $q$ (Corollary~\ref{fluxonmodcor}). In
Section~\ref{Decomp} we shall explain the connection between these two
classes of gauge theory solitons. The noncommutative instanton moduli
space coincides with the moduli space
$$
\mathcal{G}_{(p,q)}\bigl(\Torus_\Theta^2\bigr)\cong{\rm Hom}\bigl(
\pi_1(\Torus^2)\,,\,{\rm U}(N)\bigr)\,/\,{\rm U}(N)
$$
of {\it flat} principal ${\rm U}(N)$-bundles over the torus
which arises in ordinary gauge theory on
$\Torus^2$~\cite{AtiyahBott1}. In the noncommutative setting, the
moduli space is
completely determined by the symmetric product not only for flat
connections but also for {\it all} constant curvature connections. In the
commutative case, a constant curvature connection on a principal
${\rm U}(N)$-bundle over $\Torus^2$ can be described as a flat connection on
a non-trivial principal bundle over $\Torus^2$ with structure group
${\rm U}(N)\,/\,{\rm U}(1)\cong{\rm SU}(N)\,/\,\zed_N$.
\label{modsprem}\end{remark}

For the general case one has to be more careful because the distinct
unstable instantons that we have constructed in
Section~\ref{Instantons} do not represent gauge equivalence
classes~\cite{PanSz1}. From (\ref{modpqexpl}) and the reducibility of the
corresponding Weyl algebra representation there is an
isomorphism $\mod_{m\,p,m\,q}\cong\oplus^m\,\mod_{p,q}$ for any
$m\in\nat$. Both Heisenberg modules $\mod_{p,q}$ and
$\mod_{m\,p,m\,q}$ have the {\it same} constant curvature
$F_{p,q}=F_{m\,p,m\,q}$ and hence the corresponding instanton solutions
should be identified through gauge invariance. This difficulty can be
circumvented by modifying our previous definition of
partition~\cite{PanSz1,Rieffel2}.
\begin{definition}
An {\it instanton partition} of the K-theory charge
$(p,q)\in\K_0^+(\Torus_\Theta^2)$ is a collection
$\underline{\underline{(p,q)}}=\{(N_k,p_k,q_k)\}$ of triples of
integers such that
\begin{enumerate}
\item $\underline{(p,q)}=\{(N_k\,p_k,N_k\,q_k)\}$ is a partition of
  $(p,q)$;
\item $(p_k,q_k)\neq(p_l,q_l)$ for $k\neq l$; and
\item $p_k$ and $q_k$ are relatively prime for each $k$.
\end{enumerate}
\label{instpartdef}\end{definition}
\noindent
Given an arbitrary partition $\underline{(p,q)}=\{(p_k,q_k)\}$ of
$(p,q)\in\K_0^+(\Torus_\Theta^2)$, we may
write $(p_k,q_k)=N_k\,(p_k',q_k')$ for each $k$ with $N_k:={\rm
  gcd}(p_k,q_k)$ and $p_k',q_k'$ coprime. Definition~\ref{instpartdef}
then restricts to those partitions with {\it distinct} K-theory
charges $(p_k',q_k')$. They represent the distinct gauge equivalence
classes of instanton solutions to the Yang-Mills
equations~\cite{PanSz1,Rieffel2}.
\begin{theorem}
The moduli space
$\mathcal{G}_{\underline{\underline{(p,q)}}}(\Torus_\Theta^2)$ of
instantons on the Heisenberg module $\mod_{p,q}$ corresponding to an
instanton partition $\underline{\underline{(p,q)}}=\{(N_k,p_k,q_k)\}$
is given by
$$
\mathcal{G}_{\underline{\underline{(p,q)}}}\bigl(\Torus_\Theta^2\bigr)~=~
\prod_k\,{\rm Sym}^{N_k}\bigl(\tilde\Torus{}^2\bigr) \ .
$$
\label{genmodspthm}\end{theorem}
\begin{proof}
The partition $\{(N_k\,p_k,N_k\,q_k)\}$ modifies the module
decompositions used in the proof of Theorem~\ref{instsolsthm} to
$$
\mod_{p,q}=\bigoplus_k\,\mod_{N_k\,p_k,N_k\,q_k}
$$
where the constant curvatures of the submodules
$\mod_{N_k\,p_k,N_k\,q_k}$ are all distinct. Since gauge
transformations preserve the constant curvature conditions, they
also preserve each submodule $\mod_{N_k\,p_k,N_k\,q_k}$ and so the
instanton moduli space is given by the product
$$
\mathcal{G}_{\underline{\underline{(p,q)}}}\bigl(\Torus_\Theta^2\bigr)=
\prod_k\,\mathcal{G}_{(N_k\,p_k,N_k\,q_k)}\bigl(\Torus_\Theta^2\bigr) \ ,
$$
where $\mathcal{G}_{(N_k\,p_k,N_k\,q_k)}(\Torus_\Theta^2)$ is the
moduli space of constant curvature connections on the Heisenberg
module $\mod_{N_k\,p_k,N_k\,q_k}$. The result now follows by
Theorem~\ref{globalmodspthm}.
\end{proof}

\subsection{Decompactification\label{Decomp}}

The classification of classical solutions which we have given is
completely analogous to that which arises in {\it ordinary} gauge
theory on a two-dimensional torus $\Torus^2$. In the commutative case
it is just the Atiyah-Bott bundle-splitting
construction~\cite{AtiyahBott1,Griguolo1,PanSz1} and can be obtained in
the present case by formally setting $\Theta=0$ everywhere. The
integers $p$ and $p_k$ then represent the true ranks of Hermitean
vector bundles over $\Torus^2$. The difference in the noncommutative
case lies in the structure of the partitions that can contribute to
finite action solutions~\cite{PanSz1}. Furthermore, in the commutative
case if we were to {\it decompactify} the torus $\Torus^2$ onto the
plane $\plane_2$ by formally sending the modulus $\tau\to\infty$, then
the instanton solutions would ``disappear'', in the sense that their
action vanishes in this limit. This follows by setting $\Theta=0$ in
Theorem~\ref{instsolsthm}, and it is consistent with the fact that there
are no topologically non-trivial finite action gauge field
configurations on the plane. Instead, as we will now show, the formal
``decompactification'' of the noncommutative torus $\Torus_\Theta^2$
onto the Moyal plane $\plane_2^\theta$ maps the instantons of this
section in a very precise way onto the fluxons constructed in
Section~\ref{Fluxons}~\cite{GSSz1}. This
provides a rather remarkable limiting construction of the intricate
fluxon solutions given by Theorem~\ref{fluxonthm} in terms of the
relatively simpler noncommutative torus instantons obtained through
the Fock module realizations provided by
Proposition~\ref{constcurvprop}.

We introduce the quantity
\beq
\theta:=\frac{\tau\,\Theta}{2\pi}
\label{thetataudef}\eeq
which will turn out to be the noncommutativity parameter of the Moyal
plane $\plane_2^\theta$. We take the limits $\tau\to\infty$,
$\Theta\to0$ whilst keeping the combination (\ref{thetataudef})
fixed. From the form of the Yang-Mills functional given by
Theorem~\ref{instsolsthm}, the only finite action configurations which
survive this limit are those partitions
$\underline{(p,q)}=\underline{(0,-q)}_{\,0}:=\{(0,-q_k)\}$ having
$p_k=0$ and $q_k,q>0$ for all $k$ (so that $q_k\,\Theta,q\,\Theta>0$
by the partition constraints). The action of these instantons may be
written in terms of the parameter (\ref{thetataudef}) as
\beq
\YM\bigl(\,\nabla_{\underline{(0,-q)}_{\,0}}
\,\bigr)=\frac{2\pi\,q}\theta \ .
\label{YMtoruslimit}\eeq
By Theorem~\ref{fluxonthm} this is the action of a fluxon of
topological charge $q\in\nat$.

To provide the explicit mapping of instantons onto fluxons, let us fix
a finite action decompactification partition
$\underline{(0,-q)}_{\,0}$. For $p=0$ the Weyl algebra reads
$\Gamma_1\,\Gamma_2=\Gamma_2\,\Gamma_1$. With respect to the canonical
orthonormal basis $w_k$, $k=0,1,\dots,q-1$ of $\Weyl_{0,-q}$, this equation is
generically solved up to unitary isomorphism by operators of the form
\beq
\Gamma_i=\exp\bigl(\mbox{$-\frac{2\pi\ii}{\sqrt\tau}$}\,\mbf\lambda^i
\bigr):=\sum_{k=0}^{q-1}\,\e^{-\frac{2\pi\ii}{\sqrt\tau}\,\lambda_k^i}~
w_k\otimes w_k^*
\label{Gammaip0}\eeq
with $\lambda_k^i\in\real$ for $i=1,2$ and $k=0,1,\dots,q-1$. The
corresponding module splitting provided by Theorem~\ref{instsolsthm}
thereby yields an isomorphism
\beq
\mod_{0,-q}=\bigoplus_{k=0}^{q-1}\,\mod_{0,-q_k}\cong\Fock^q \ .
\label{mod0qisoFock}\eeq

The global minimum of the Yang-Mills functional on
(\ref{mod0qisoFock}) is the connection with constant curvature
\beq
F_{0,-q}=\mbox{$\frac\ii\theta$}~\P_{0,-q} \ .
\label{F0qconst}\eeq
The projector $\P_{0,-q}$ has rank $\Tr(\P_{0,-q})=q\,\Theta$, so that
$\frac1\Theta\,\P_{0,-q}$ has integer rank $q\in\nat$. By representing
the corresponding K-theory class by the Boca projection one can show
that the image of $\frac1\Theta\,\P_{0,-q}$ is a
$q$-dimensional subspace of the Fock module (\ref{mod0qisoFock}) in
the decompactification limit, and thus the curvature (\ref{F0qconst})
coincides with that of the $q$-fluxon solution of
Theorem~\ref{fluxonthm}~\cite{KrajSch1,LLSz2}.
\begin{proposition}
As endomorphisms of the Fock module $\Fock$ one has
$$
\lim_{\Theta\to0}\,\mbox{$\frac1\Theta$}\,\P_{0,-q}=\P_{(q)}=
\sum_{k=0}^{q-1}\,e_k\otimes e_k^* \ .
$$
\label{projlimlemma}\end{proposition}

In the decompactification limit, we would like to map the
noncommutative torus algebra $\Torus_\Theta^2$ onto the Moyal algebra
$\plane_2^\theta$ at the level of their representations as
endomorphisms of the Fock module. At the level of generators this
means that we would
like to roughly identify $U_i$, acting on (\ref{mod0qisoFock}) via
(\ref{Uimodpq}), with the exponential operators
$\exp\bigl(\frac{2\pi\ii}{\sqrt\tau}\,x^i\bigr)$ where the Moyal plane
generators $x^i$ obey $[x^1,x^2]=\ii\theta\,\one$ via the
Baker-Campbell-Hausdorff formula. The immediate problem which arises
is that the $x^i$ are defined on the trivial rank~$1$ module
$\Hil=\plane_2^\theta$, while
\beq
U_i=\exp\bigl(-\mbox{$\frac{2\pi}{\sqrt\tau}$}\,(\theta\,
\nabla_i^{\rm c}\otimes\one_q+\ii\one\otimes\mbf\lambda^i)\bigr)
\label{UiFock}\eeq
are defined on the Fock module $\Fock^q$. In order to make an
identification of this type we need to ensure that all operators are
defined on a common domain.

We first embed all the pertinent operators naturally into the
projective $\plane_2^\theta$-module $\mod_{0,-q}\oplus\Hil$ via the
definitions
\bea
\hat\nabla_i^{\rm c}&=&\bigl(\nabla_i^{\rm c}\otimes\one_q\bigr)
{}~\oplus~0 \ , \nonumber\\[4pt] \hat{\mbf\lambda}{}^i&=&
\bigl(\one\otimes\mbf\lambda^i\bigr)~\oplus~0 \ , \nonumber\\[4pt]
\hat x^i&=&\mbf0_q~\oplus~x^i \ .
\label{trivembed}\eea
Then we represent these endomorphisms on the free module $\Hil$ by
finding their images under a natural unitary isomorphism of separable
Hilbert spaces
\beq
\Xi_q\,:\,\mod_{0,-q}\oplus\Hil~\stackrel{\approx}
{\longrightarrow}~\Hil \ .
\label{Omegaqiso}\eeq
To construct this isomorphism, let $\shift$ be the shift
endomorphism of the Fock module with $\ker(\shift)^q=\{0\}$ and
$\ker(\shift^\dag)^q={\rm im}(\P_{(q)})\cong\complex^q$. Then the
submodule $(\shift)^q\cdot\Hil$ is the orthogonal complement in $\Hil$
of $\mod_{0,-q}\cong\P_{(q)}\cdot\Hil$. The isomorphism
(\ref{Omegaqiso}) is therefore given explicitly by
\beq
\Xi_q\bigl(\,\mbox{$\sum\limits_{k=0}^{q-1}$}\,f\cdot
e_k~\oplus~f'\,\bigr):=\P_{(q)}\cdot f+(\shift)^q\cdot f'
\label{Omegaqdef}\eeq
for $f,f'\in\Hil$. The inverse map is given for $f\in\Hil$ by
\beq
\Xi_q^{-1}(f)=\sum_{k=0}^{q-1}\,f\cdot e_k~\oplus~\bigl(\shift^\dag
\bigr)^q\cdot f \ .
\label{Omegaqinv}\eeq

It is straightforward to work out the action of the isomorphism
(\ref{Omegaqiso}) on the operators (\ref{trivembed}) and as
endomorphisms of $\Hil$ one finds
\beq
\Xi_q\,\hat x^i\,\Xi_q^{-1}&=&\bigl(\shift\bigr)^q\,x^i\,
\bigl(\shift^\dag\bigr)^q \ , \nonumber\\[4pt]
\Xi_q\,\hat{\mbf\lambda}{}^i\,\Xi_q^{-1}&=&
\sum_{k=0}^{q-1}\,\lambda_k^i~e_k\otimes e_k^* \ .
\label{Omegaqxlambda}\eea
The desired identifications are now given by
\beq
U_i=\Xi_q\,\exp\bigl(-\mbox{$\frac{2\pi}{\sqrt\tau}$}\,(
\theta\,\hat\nabla_i^{\rm c}+\ii\hat{\mbf\lambda}{}^i)\bigr)\,\Xi_q^{-1}
:=\Xi_q\,\exp\bigl(\mbox{$\frac{2\pi\ii}{\sqrt\tau}$}\,
\hat x^i\bigr)\,\Xi_q^{-1}
\label{Uixiid}\eeq
for $i=1,2$. From (\ref{Omegaqxlambda}) it then follows finally that
the decompactification limit of the constant curvature connection of
Yang-Mills theory on $\Torus_\Theta^2$ thus leads to the operators
\beq
D_i:=\ii\theta\,\Xi_q\,\hat\nabla_i^{\rm c}\,\Xi_q^{-1}=
\sum_{k=0}^{q-1}\,\lambda_k^i~e_k\otimes e_k^*+
\bigl(\shift\bigr)^q\,x^i\,\bigl(\shift^\dag\bigr)^q \ .
\label{Didecomp}\eeq
This coincides with the $q$-fluxon given by
Theorem~\ref{fluxonthm}. Thus the decompactification of finite action
instantons on the noncommutative torus provides a natural and clear
way to describe the fluxon solutions of Yang-Mills gauge theory on
$\plane_2^\theta$. The construction of these solutions is very natural
in the toroidal framework, and the torus instanton origin of fluxons
explains many of their seemingly unusual properties in precise
geometric ways~\cite{GSSz1}. For instance, the constraint $q>0$ on the
sign of the fluxon charges can be traced back to the required positivity of the
Murray-von~Neumann dimension on the positive K-theory cone of stable
Heisenberg modules, while the instability of fluxons is due to the
instability of their instanton ancestors.

\newsection{D-branes in group manifolds\label{GroupD}}

In this final section we leave the setting of flat target spaces $X$
and study some examples of D-branes in curved backgrounds. A
particularly tractable class of examples is provided by the cases
where $X={\rm G}$ is a {\it group} manifold. While these spacetimes are not
entirely realistic string backgrounds, they provide important solvable
models which sometimes form subspaces of genuine target spaces. They
possess enough symmetries so that a relatively complete classification
of D-branes, and their noncommutative worldvolume geometries, may be readily
obtained. We will describe the general quantization scheme for special
classes of D-branes in these backgrounds, and how to construct the
corresponding noncommutative worldvolume gauge theories. We then work
out the simplest example of D-branes whose worldvolumes are fuzzy
two-spheres in the group manifold of ${\rm G}={\rm SU}(2)$ where
everything can be made very explicit. A detailed review of these
matters along with an exhaustive list of references can be found
in~\cite{Schomrev1}.

\subsection{Symmetric D-branes\label{SymDbranes}}

Let ${\rm G}$ be a compact, simple, simply connected and connected Lie group
possessing a bi-invariant metric. Let $\mfg$ be the Lie algebra of
${\rm G}$. The Lie bracket on $\mfg$ is denoted
$[-,-]:\mfg\times\mfg\to\mfg$, and the metric on ${\rm G}$ induces
an invariant inner product
$\langle-,-\rangle:\mfg\times\mfg\to\real$. As in Section~\ref{Intro},
a string in the target space $X={\rm G}$ is a harmonic map
$g:\Sigma\to {\rm G}$ on an oriented Riemann surface $\Sigma$. The
special feature of group manifolds is that the string theory possesses
an affine ${\rm G}\times\overline{{\rm G}}$ symmetry
\beq
g(z,\bar z)~\longmapsto~\Omega(z)\,g(z,\bar z)\,\overline{\Omega}(\bar
z)^{-1} \ ,
\label{GbarGsym}\eeq
where $(z,\bar z)$ are local complex coordinates on $\Sigma$ and
$\Omega,\overline{\Omega}:\Sigma\to {\rm G}$ are independent holomorphic and
antiholomorphic maps.

Select a fixed boundary component of $\partial\Sigma$, and choose the
local parametrization of the worldsheet $\Sigma$ such that this component is
located at $z=\bar z$. Then the symmetry (\ref{GbarGsym}) restricts to
this boundary component as
\beq
g\big|_{\partial\Sigma}~\longmapsto~\Omega\,g\big|_{\partial\Sigma}\,
\overline{\Omega}{}^{\,-1} \ .
\label{Gsymbdry}\eeq
A D-brane is now a boundary condition
$g\big|_{\partial\Sigma}:\partial\Sigma\to W\subset {\rm G}$ which preserves
enough of the affine ${\rm G}\times\overline{{\rm G}}$ symmetry as
dictated by conformal invariance of the underlying boundary conformal
field theory.
\begin{definition}
A {\it symmetric D-brane} in ${\rm G}$ is a boundary condition with
$$
\overline{\Omega}=\omega\circ\Omega
$$
on $\partial\Sigma$ for some isometric automorphism $\omega:{\rm G}\to
{\rm G}$.
\label{symDbranedef}\end{definition}

Symmetric D-branes preserve a maximal diagonal subgroup ${\rm G}\subset
{\rm G}\times\overline{{\rm G}}$ of the affine symmetry. Their worldvolumes
$W$ are {\it twisted conjugacy classes}
\beq
W={\rm C}_\omega(g)=\bigl\{h\,g\,\omega(h)^{-1}~\big|~h\in {\rm G}\bigr\}
\label{twistconjclass}\eeq
of elements $g\in {\rm
  G}$~\cite{AlekSch1,FigStan1,Stanciu1,FroTwist1,FluxStab1}.
\begin{definition}
Two symmetric D-branes ${\rm C}_\omega(g)$ and ${\rm C}_\omega(g'\,)$ are {\it
  equivalent} if $g'={\rm Ad}_h(g):=h\,g\,h^{-1}$ for some $h\in {\rm G}$.
\label{equivDbranedef}\end{definition}
\noindent
Equivalent D-branes are described by twisted conjugacy classes which
are simply translates of one another in the group manifold of ${\rm G}$. To
characterize the corresponding equivalence classes, let ${\rm Aut}({\rm G})$
denote the group of isometric automorphisms of ${\rm G}$. Let ${\rm
  Inn}({\rm G})\subset{\rm Aut}({\rm G})$ be the invariant normal subgroup of
inner automorphisms $g\mapsto{\rm Ad}_h(g)$. Then the factor group
\beq
{\rm Out}({\rm G})={\rm Aut}({\rm G})\,/\,{\rm Inn}({\rm G})
\label{OutGdef}\eeq
consists of equivalence classes of metric-preserving outer
automorphisms of ${\rm G}$. To each element of the group (\ref{OutGdef}) we
can associate an equivalence class of symmetric D-branes foliating
${\rm G}$, because every element of ${\rm G}$ belongs to one and only
one twisted conjugacy class. The leaves of this foliation need not all
have the same topology. Hence the D-brane foliation need not be a
fibration.

Thus a generic worldvolume (\ref{twistconjclass}) representing an
equivalence class of D-branes is described by
taking $\omega\in{\rm Out}({\rm G})$ to be an outer automorphism. The
symmetric D-brane is then called an {\it $\omega$-twisted
  D-brane}. When $\omega={\rm id}_{\rm G}$ then ${\rm C}(g):={\rm
  C}_{{\rm id}_{\rm G}}(g)$ is
just an ordinary {\it conjugacy class} of the group ${\rm G}$ and the
symmetric D-brane is called an {\it untwisted D-brane}. Generally, the
twisted conjugacy class is diffeomorphic to a homogeneous space
\beq
{\rm C}_\omega(g)={\rm G}\,/\,{\rm H}_\omega(g)
\label{twistconjhomsp}\eeq
where
\beq
{\rm H}_\omega(g)=\bigl\{h\in{\rm G}~\big|~h\,g=g\,\omega(h)\bigr\}
\label{isotropysubgp}\eeq
is the isotropy subgroup of the element $g\in {\rm G}$.
\begin{remark}
It is possible to also construct D-branes with less symmetry,
preserving a smaller subgroup than the diagonal ${\rm G}\subset{\rm
  G}\times\overline{\rm G}$. These are called {\it symmetry-breaking
  D-branes}. Generically, they are localized along {\it products} of
twisted conjugacy classes of the Lie group $\rm G$~\cite{Quella1}.
\label{symbreakrem}\end{remark}

The supergravity fields on a symmetric D-brane are straightforward to
construct~\cite{AlekSch1,Stanciu1,FluxStab1}. For a given group
element $g\in{\rm G}$, identify the tangent space
$T_g{\rm G}\cong\mfg$ to $\rm G$ at $g$ with the Lie algebra of $\rm G$. For
tangent vectors $u,v\in\mfg$ the invariant metric $G$ is then given by
\beq
G(u,v)=\bigl\langle g^{-1}\,u\,,\,g^{-1}\,v\bigr\rangle \ ,
\label{metricG}\eeq
while the invariant three-form $H$ is
\beq
H(u,v,w)=-\bigl\langle g^{-1}\,u\,,\,\bigl[g^{-1}\,v\,,\,
g^{-1}\,w\bigr]\bigr\rangle
\label{3formG}\eeq
for $u,v,w\in\mfg$. We may introduce a $B$-field, with $H=\dd
B$, by the formula
\beq
B(u,v)=\bigl\langle g^{-1}\,u\,,\,\frac{\one+\omega\circ{\rm Ad}_g}
{\one-\omega\circ{\rm Ad}_g}\bigl(g^{-1}\,v\bigr)\bigr\rangle
\label{BfieldG}\eeq
which is defined for $g^{-1}\,u\in{\rm im}(\one-\omega\circ{\rm
  Ad}_g)$, i.e. for vectors $u$ tangent to the twisted conjugacy class
(\ref{twistconjclass}) containing $g\in{\rm G}$. We assume that these
fields restrict non-degenerately to the twisted conjugacy classes.

\subsection{Untwisted D-branes\label{Untwist}}

Let us first describe the noncommutative worldvolume geometries in the
somewhat more standard cases with
$\omega={\rm id}_{\rm G}$. The D-brane worldvolumes are the conjugacy
classes $W={\rm C}(g)=\{{\rm Ad}_h(g)~|~h\in{\rm G}\}$ which are
diffeomorphic to the symmetric spaces ${\rm G}\,/\,{\rm H}(g)$, where
${\rm H}(g)$ is the stabilizer subgroup of the element $g\in{\rm
  G}$. Since $\rm G$ is compact and simple, every element is conjugate
to some maximal torus $\Torus$ of $\rm G$. Let us restrict to the set
${\rm G}_{\rm r}$ of {\it regular} elements $g\in{\rm G}$, i.e. those elements
which are conjugate to only one maximal torus. The set ${\rm G}_{\rm r}$
is an open dense subset in $\rm G$. Let $\Torus_{\rm
  r}:=\Torus\cap{\rm G}_{\rm r}$. Then ${\rm
  C}(g)\cap\Torus_{\rm r}\neq\emptyset$ and the intersections generate
an orbit of the Weyl group ${\rm W}={\rm N}(\Torus_{\rm
  r})\,/\,\Torus_{\rm r}$, where ${\rm N}(\Torus_{\rm r})$ is the
normalizer subgroup of the maximal torus. Hence there is a
diffeomorphism realizing the conjugacy class as the flag manifold
\beq
{\rm C}(g)={\rm G}_{\rm r}\,/\,\Torus_{\rm r} \ .
\label{Cgflagmfld}\eeq
Since the Weyl group $\rm W$ further relates elements of $\Torus_{\rm r}$,
it follows that untwisted D-branes are parametrized by the Weyl
chamber $\Torus_{\rm r}\,/\,{\rm W}$.

Our main result here is that the untwisted D-branes in $\rm
G$ in the semi-classical regime recover the Kirillov theory of
coadjoint orbits~\cite{ARS1}, whose quantization gives all irreducible
representations of the universal enveloping algebra $U(\mfg)$.
\begin{theorem}
To each irreducible representation $\mathcal{V}$ of the Lie group
$\rm G$ there bijectively corresponds an untwisted D-brane in the
semi-classical limit such that the noncommutative algebra of functions
on the quantized worldvolume $W_{\mathcal{V}}$ is given by
$$
\alg(W_{\mathcal{V}})~=~{\rm End}(\mathcal{V}) \ .
$$
\label{NCalguntwist}\end{theorem}
\begin{proof}
The two-form $B$-field, defined on vectors tangent to the conjugacy
class ${\rm C}(g)$, is given by the automorphism
$$
B=\frac{\one+{\rm Ad}_g}{\one-{\rm Ad}_g} \ .
$$
In the semi-classical limit, $H\to0$ and the group manifold of $\rm G$
approaches flat space. We may then parametrize the conjugacy class by
an element $X\in\mfg$ such that $g\approx\one+X$ in the limit. The
semi-classical $B$-field thereby becomes
$$
B=-2\,({\rm ad}_X)^{-1} \ ,
$$
which is just the standard Kirillov two-form on the Lie algebra
$\mfg$. The Seiberg-Witten bivector is given by
$$
\theta=\frac2{B-G\,B^{-1}\,G}=\mbox{$\frac12$}\,\bigl({\rm Ad}_{g^{-1}}-
{\rm Ad}_g\bigr) \ .
$$
In the limit $H\to0$ this bivector obeys the Jacobi identity and
becomes
$$
\theta={\rm ad}_X \ ,
$$
which is the Kirillov-Kostant Poisson bivector. Casimir operators in
$U(\mfg)$ are invariants of the conjugacy classes and hence may be
used to label ${\rm C}(g)$. We may therefore
associate to each conjugacy class ${\rm C}(g)$ whose second Casimir
invariant is quantized in the requisite way an irreducible
representation $\mathcal{V}_g$ of $U(\mfg)$.
\end{proof}
\begin{remark}
This theorem is consistent with the fact that conjugacy classes are
the image under the exponential map of adjoint orbits, which are
diffeomorphic to coadjoint orbits that are symplectic manifolds with
respect to the natural Kirillov-Kostant-Souariu symplectic structure
and hence are even-dimensional. Note that the invariant inner product
$\langle-,-\rangle$ on $\mfg$ plays a crucial role in this
identification. It relates the D-brane worldvolume, which is an orbit
of the {\it adjoint} action of the group $\rm G$ on itself, to the
{\it coadjoint} action of $\rm G$ on its Lie algebra $\mfg$. The
quantization of the coadjoint orbits in turn gives the representations
of $\rm G$.
\label{KKSsymplrem}\end{remark}
\begin{remark}
The proof of Theorem~\ref{NCalguntwist} shows that the semi-classical
geometry of untwisted D-branes is very close to that of the cases studied in
earlier sections with constant $B$-field~\cite{ARS1}. Since $\rm G$ is
compact, the algebra $\alg(W_{\mathcal{V}})$ is finite-dimensional and
so the noncommutative worldvolume geometry is now ``fuzzy''. This
algebra carries a natural $\rm G$-action on it which represents the
isometry group of the noncommutative space. Note that not all conjugacy
classes are admissible as D-brane worldvolumes. Only the {\it integer}
conjugacy classes which are in one-to-one correspondence with the
irreducible representations of $\rm G$ are allowed. The algebra of
functions on the quantized conjugacy classes is then the corresponding
endomorphism algebra of the representation.
\label{fuzzyrem}\end{remark}
\begin{remark}
In some instances Theorem~\ref{NCalguntwist} also extends to the case
of non-compact and even non-semisimple Lie groups $\rm G$,
such as those corresponding to homogeneous plane wave
backgrounds~\cite{SamSz1}. It is not clear how to generalize this
result to symmetry-breaking D-branes whose classical worldvolumes are
given as products of conjugacy classes~\cite{Quella1}.
\label{genfuzzyrem}\end{remark}

\subsection{Example: D-branes in SU(2)\label{DbranesS3}}

Let us now study in detail the simplest example of this
construction~\cite{ARS1,ARS2,Schomrev1}. The rank~$1$ Lie group ${\rm
  G}={\rm SU}(2)$ has no
non-trivial Dynkin diagram automorphisms and hence all symmetric
D-branes in ${\rm SU}(2)\cong\S^3$ are untwisted. In the
semi-classical limit $H\to0$, the radius of the three-sphere $\S^3$
grows and the group manifold ``decompactifies'' onto flat space
$\real^3={\rm su}(2)$. The conjugacy classes are parametrized by the
Weyl chamber $\S^1\,/\,\zed_2$. Let $\vartheta\in[0,\pi]$ be the coordinate
on this closed interval. Then $\vartheta$ parametrizes the azimuthal angle
of $\S^3$. For $\vartheta=0,\pi$, the conjugacy classes are points
corresponding to elements in the center $\zed_2=\{-\one,\one\}$ of
${\rm SU}(2)$ placed at the north and south poles of the
three-sphere. They correspond to D0-branes. For $\vartheta\in(0,\pi)$
the conjugacy classes are diffeomorphic to two-spheres
$\S^2\cong\S^3\,/\,\S^1$, corresponding to D2-branes. The group
manifold $\S^3=\real^3\cup\{\infty\}$ has a standard foliation by
two-spheres of increasing radius with two degenerate spheres $\S^2$
placed at $0$ and $\infty$. Because of the degeneration of the
limiting spheres, the foliation is not a fibration (In particular this
is {\it not} the Hopf fibration $\S^3\to\S^2$ that we are describing
here).

In this case $\theta={\rm ad}_X$ is the standard Kirillov-Kostant
Poisson bivector on the two-spheres in ${\rm su}(2)=\real^3$, and
the worldvolume algebras of D-branes in ${\rm SU}(2)$ are the usual
quantizations of these two-spheres via the coadjoint orbit
method. By Theorem~\ref{NCalguntwist}, quantizing functions on $\S^2$
with the usual Poisson structure yields {\it fuzzy
  spheres}~\cite{Madore1}
\beq
\alg\bigl(\S_j^2\bigr)=\mat_N\bigl(\complex\bigr)
\label{fuzzyS2def}\eeq
which are labelled by half-integers $j\in\frac12\,\nat_0$. The D-brane
label $j$ is proportional to the radius of its worldvolume $\S^2$ and
it represents the spin of the associated irreducible ${\rm
  SU}(2)$-module $\mathcal{V}_j$ of dimension $N=2j+1$. The radii $R_j$
of the corresponding integer conjugacy classes in ${\rm SU}(2)$ are
quantized as $R_j^2=j\,(j+1)$ in order to match the corresponding
second Casimir invariants.

The worldvolume algebra (\ref{fuzzyS2def}) has finite dimension
$(2j+1)^2$. It is thus a full matrix algebra
which admits an action of ${\rm SU}(2)$ by conjugation with group
elements in the $N$-dimensional representation of ${\rm SU}(2)$. Under
this action, the ${\rm SU}(2)$-module $\mat_N(\complex)$ decomposes
into a direct sum of irreducible representations $\mathcal{V}_J$ of
dimension $2J+1$, giving~\cite{Madore1}
\beq
\alg\bigl(\S_j^2\bigr)=\bigoplus_{J=0}^{N-1}\,\mathcal{V}_J \ .
\label{fuzzyS2decomp}\eeq
Note that only integer values of the spin $J$ appear in this direct
sum. Let $Y_a^J$, $a=1,\dots,2J+1$ be a basis of the representation
space $\mathcal{V}_J$. These elements are called {\it fuzzy spherical
harmonics}
and their multiplication rules can be worked out from the
multiplication of $N\times N$ matrices to get~\cite{Madore1}
\beq
Y_a^I\,Y_b^J=\sum_{K=0}^{\min(I+J,2j)}~\sum_{c=1}^{2K+1}\,\left[\,
{}^I_a\,{}^J_b\,{}^K_c\,\right]\,\left\{\,{}^I_j\,{}^J_j\,
{}^K_j\,\right\}~Y_c^K \ ,
\label{Ymultrules}\eeq
where the square brackets denote the Clebsch-Gordan coefficients of
${\rm SU}(2)$ and the curly brackets are the Wigner $6j$-symbols of
$U({\rm su}(2))$.

To construct field theories on the noncommutative worldvolumes
(\ref{fuzzyS2def}), we will
introduce derivations on $\alg(\S_j^2)$ in analogy with the flat space case in
Section~\ref{DerivMoyal} by finding automorphisms of the
noncommutative algebra which represent isometries of the
sphere $\S^2$. Recall that the classical ${\rm su}(2)$ symmetry on the
commutative algebra of functions $C(\S^2)$ is inherited from
infinitesimal rotations in $\real^3$ as follows. Let $y=(y^i)_{i=1,2,3}$
denote local coordinates on $\real^3$. Let $C^{ijk}$ be the
structure constants of the Lie algebra ${\rm su}(2)$ in a suitable
basis. Then
\beq
L_i=\sum_{j,k=1}^3\,C^{ijk}\,y^j\mbox{$\frac\partial{\partial y^k}$}
\ , \quad i=1,2,3
\label{Livecfields}\eeq
generate an ${\rm su}(2)$ subalgebra in the Lie algebra of vector
fields on $\real^3$. The classical functions $Y_a^J$ above are then
the usual spherical harmonics transforming under $L_i$ in the representation
$\mathcal{V}_J$ of ${\rm su}(2)$. Under the embedding
$\S^2\subset\real^3$ defined by $|y|^2=1$, this descends to an action
of ${\rm SU}(2)$ on the algebra $C(\S^2)$.

Let us now write down the quantization of this ${\rm su}(2)$
symmetry. Let $f=\sum_{J,a}\,f_{Ja}~Y_{a}^J$ be an element of the
worldvolume algebra (\ref{fuzzyS2decomp}). In analogy with the flat
space case, we then introduce derivatives through the adjoint
representation of ${\rm su}(2)$ by
\beq
L_i(f):=\bigl[Y_i^1\,,\,f\bigr] \ , \quad i=1,2,3 \ .
\label{S2derivdef}\eeq
The automorphisms $L_i:\alg(\S_j^2)\to\alg(\S_j^2)$ generate the
reducible action of ${\rm SU}(2)$ on the noncommutative worldvolume
algebra under which it decomposes as in eq.~(\ref{fuzzyS2decomp}).

We are now ready to describe gauge theory on the noncommutative space
(\ref{fuzzyS2def}). As always, the standard procedure for constructing
connections and gauge theories in noncommutative geometry can be
formally developed in this instance~\cite{Madore1}. Here we will just
write down the final result for a rank~$1$ gauge theory on the
trivial projective module over the fuzzy sphere. We express an arbitrary
connection $\nabla_i:\alg(\S_j^2)\to\alg(\S_j^2)$ in this case in the
form
\beq
\nabla_i=L_i+A_i
\label{connfuzzyS2}\eeq
with $A_i\in\ii{\rm u}(N)$, $i=1,2,3$. Unlike the previous flat space
actions,
string theory considerations~\cite{ARS2} dictate that the gauge theory action
functional $S:\ii{\rm u}(N)\to\real$ contains both Yang-Mills and
Chern-Simons terms in the combination
\beq
S(A)=\Tr\Bigl(\,\sum_{i,j=1}^3\,\biggl[\bigl(F_{ij}\bigr)^2+2\,
\sum_{k=1}^3\,C^{ijk}~{\rm CS}_{ijk}(A)\biggr]\,\Bigr) \ .
\label{fuzzyS2action}\eeq
Here $\Tr$ denotes the usual $N\times N$ matrix trace on
$\mat_N(\complex)$, and
\beq
F_{ij}:=\bigl[\nabla_i\,,\,\nabla_j\bigr]=
L_i(A_j)-L_j(A_i)+\bigl[A_i\,,\,A_j\bigr]-\ii
\sum_{k=1}^3\,C^{ijk}\,A_k
\label{curvconnfuzzyS2}\eeq
is the curvature of the connection (\ref{connfuzzyS2}). The functional
\beq
{\rm CS}_{ijk}(A)=\ii L_i(A_j)\,A_k+\mbox{$\frac13$}\,A_i\,\bigl[A_j\,,\,
A_k\bigr]+\mbox{$\frac12$}\,\sum_{l=1}^3\,C^{ijl}\,A_l\,A_k
\label{NCCSform}\eeq
is the {\it noncommutative Chern-Simons form}~\cite{ARS2}. The action
functional (\ref{fuzzyS2action}) is invariant under the
(infinitesimal) gauge transformations
\beq
A_i~\longmapsto~A_i+L_i(\Lambda)-\ii\bigl[A_i\,,\,\Lambda\bigr]
\label{fuzzyS2gaugetransfs}\eeq
with $\Lambda\in{\rm u}(N)$.

Varying the functional (\ref{fuzzyS2action}) gives the equations of
motion
\beq
\sum_{i=1}^3\,\left(L_i(F_{ij})+\bigl[A_i\,,\,F_{ij}\bigr]\right)=0 \
, \quad j=1,2,3
\label{fuzzyS2eom}\eeq
expressing the usual fact that the curvature is covariantly
constant. To solve these equations, introduce the shifted algebra
elements
\beq
R_i:=Y_i^1+A_i
\label{BifuzzyS2def}\eeq
in $\mat_N(\complex)$ to write them as
\beq
\sum_{i=1}^3\,\bigl[R_i\,,\,\bigl[R_i\,,\,R_j\bigr]-
\mbox{$\sum\limits_{k=1}^3$}\,C^{ijk}\,R_k\bigr]=0 \ .
\label{eomBi}\eeq
There are then two classes of solutions. The first class arises from
requiring that all three matrices (\ref{BifuzzyS2def}) be mutually
commuting in $\mat_N(\complex)$, $[R_i,R_j]=0$ for $i,j=1,2,3$. They
can therefore be simultaneously diagonalized and their simultaneous eigenvalues
describe translates of $N$ particles in the group target space $X={\rm
  SU}(2)$. These are formally the same as the solutions found earlier
in the cases of flat spaces~\cite{Wittenp}.

A more interesting class of solutions with no flat space analog is
provided by those configurations (\ref{BifuzzyS2def}) which obey the
commutation relations
\beq
\bigl[R_i\,,\,R_j\bigr]=\sum_{k=1}^3\,C^{ijk}\,R_k \ .
\label{BiSU2rep}\eeq
Such solutions have vanishing curvature $F_{ij}=0$ and thus correspond
to flat connections on the D-brane worldvolume. They determine
$N$-dimensional unitary representations of ${\rm su}(2)$,
i.e.~homomorphisms
\beq
\pi_N\,:\,{\rm su}(2)~\longrightarrow~{\rm u}(N) \ .
\label{piNhomo}\eeq
Up to isomorphism, for any $n\in\nat$ there is a unique irreducible
representation of ${\rm SU}(2)$ of dimension $n$. Thus to any
representation (\ref{piNhomo}) we can assign an unordered {\it
  partition} $(n_i)_{i=1,\dots,r}$ of $N=n_1+\dots+n_r$, with $n_i$
giving the dimensions of the irreducible submodules in
$\pi_N$. These configurations are thus similar to the torus instantons
that we constructed in Section~\ref{Instantons}. The partition
characterizes the original representation uniquely up to gauge
equivalence, and hence provides a simple classification for solutions
of this type. It is also possible to formulate gauge theory on the fuzzy
sphere in such a way that the classical solution set resembles more
closely that of Yang-Mills theory on $\plane_2^\theta$ and
$\Torus_\Theta^2$, with intimate relationships between the
seemingly distinct instanton configurations on the diverse
noncommutative spaces~\cite{Stein1,SteinSz1}.
\begin{remark}
In the string theory setting, these solutions describe dynamical
processes involving a stack of $N$ particles corresponding to D0-branes,
which are labelled by quantum mechanical instanton-type partition
degrees of freedom. These D0-branes ``decay'' into a single D2-brane
(for rank~$1$ gauge theory) with spherical worldvolume $\S^2$
corresponding to the irreducible representation $\mathcal{V}_j$ of dimension
$N=2j+1$~\cite{ARS2}. This condensation phenomenon is called the {\it
  dielectric effect}~\cite{Myers1} and it is equivalent to vector
bundle modification when D-branes are regarded as Baum-Douglas
K-cycles in topological K-homology~\cite{HM1,AST1,RSz1,Sz1}.
\label{fuzzyS2Dbrane}\end{remark}

\subsection{Twisted D-branes\label{Twist}}

We close this final section with a tour beyond the ${\rm SU}(2)$
target space and untwisted D-branes. Let us consider the generic case
of an $\omega$-twisted D-brane corresponding to a non-trivial outer
automorphism $\omega$ of the Lie group $\rm G$. In this case, the
dimension of a twisted conjugacy class ${\rm C}_\omega(g)$ is larger
than the dimension of a regular conjugacy class. In particular, there
are generically instances in which ${\rm C}_\omega(g)$ is
odd-dimensional, and so in general the worldvolume $W$ will not be a
symplectic manifold. Nevertheless, there is a way to quantize these
geometries that we shall now describe.

The main result here is that the $\omega$-twisted D-branes in $\rm G$
are labelled by representations of the {\it invariant subgroup}
\beq
{\rm G}^\omega:=\bigl\{g\in{\rm G}~\big|~\omega(g)=g\bigr\}
\label{Gomegadef}\eeq
which for $\omega\neq{\rm id}_{\rm G}$ is a proper subgroup of $\rm
G$. By conjugating to the maximal torus $\Torus$, they are thus
parametrized by the abelian subgroup $\Torus^\omega:={\rm
  G}^\omega\cap\Torus$ and there is a diffeomorphism
\beq
{\rm C}_\omega(g)={\rm G}\,/\,\Torus^\omega
\label{Comegagflag}\eeq
for any $g\in{\rm G}$. In the semi-classical regime, the quantization
of twisted conjugacy classes, i.e. the noncommutative geometry of
twisted D-branes, can be described as follows~\cite{AFQS1}.
\begin{theorem}
To each irreducible representation $\mathcal{V}^\omega$ of the invariant
subgroup ${\rm G}^\omega$ there bijectively corresponds an
$\omega$-twisted D-brane in the semi-classical limit such that the
noncommutative algebra of functions on the quantized worldvolume
$W_{\mathcal{V}^\omega}$ is given by
$$
\alg\bigl(W_{\mathcal{V}^\omega}\bigr)~=~\bigl(C({\rm G})\otimes
{\rm End}(\mathcal{V}^\omega)\bigr)^{{\rm G}^\omega} \ ,
$$
where the superscript denotes the ${\rm G}^\omega$-invariant part and
the ${\rm G}^\omega$-action ${\rm G}^\omega\times C({\rm G},{\rm
  End}\,\mathcal{V}^\omega)\to C({\rm G},{\rm
  End}\,\mathcal{V}^\omega)$ is defined by $
(h,f(g))\mapsto\mathcal{V}^\omega(h)\,f(g\,h)\,
\mathcal{V}^\omega(h)^{-1}$
with $\mathcal{V}^\omega(h)\in{\rm GL}(\mathcal{V}^\omega)$ for all
$h\in{\rm G}^\omega$.
\label{twistedthm}\end{theorem}
\begin{proof}
Consider an open neighbourhood $U$ of the identity element $\one$ of
$\rm G$. Let $g\in U$. Then the twisted conjugacy class of $g$ can be
represented as the fibration
$$
{\rm C}_\omega(g)={\rm G}\times_{{\rm G}^\omega}{\rm C}'(g)
$$
over ${\rm G}\,/\,{\rm G}^\omega$ with fiber ${\rm C}'(g)$ which is a
regular conjugacy class of the invariant subgroup ${\rm
  G}^\omega$. Here ${\rm G}^\omega$ acts on ${\rm G}$ by right
multiplication. As before in the untwisted case, in the semi-classical
limit $H\to0$ the conjugacy class ${\rm C}'(g)$ becomes small and
approaches a coadjoint orbit of~${\rm G}^\omega$, while the
Poisson manifold ${\rm G}\,/\,{\rm G}^\omega$ grows (approaching flat
space) and its Poisson bivector scales down. Thus ${\rm C}'(g)$
becomes a noncommutative symplectic space while ${\rm G}\,/\,{\rm
  G}^\omega$ remains a classical space in the semi-classical
regime. After quantization, we get a bundle with noncommutative fibers
${\rm End}(\mathcal{V}_g^\omega)$ and a classical base ${\rm
  G}\,/\,{\rm G}^\omega$.
\end{proof}
\begin{remark}
$\alg(W_{\mathcal{V}^\omega})$ is an associative matrix algebra of
functions on the Lie group $\rm G$. If
$\mathcal{V}^\omega\cong\complex$ is the trivial representation of
${\rm G}^\omega$, then the noncommutative algebra $\alg(W_\complex)$
consists of functions on $\rm G$ which are simply invariant under
right translations by elements of the invariant subgroup ${\rm
  G}^\omega\subset{\rm G}$. On the other hand, when $\omega={\rm
  id}_{\rm G}$ is the trivial automorphism of $\rm G$, one has ${\rm
  G}^\omega={\rm G}$ and the noncommutative worldvolume algebra is
${\rm End}(\mathcal{V}^{{\rm id}_{\rm G}})$ consistently with
Theorem~\ref{NCalguntwist}.
\label{twistedalgrem}\end{remark}

Using the fact that the Lie group $\rm G$ is simple, simply-connected
and compact, we can obtain an alternative realization of the
noncommutative worldvolume algebra $\alg(W_{\mathcal{V}^\omega})$
which makes its $\rm G$-module structure more transparent.
\begin{theorem}
All complexified vector bundles ${\rm G}\times_{{\rm
    G}^\omega}(\mathcal{V}^\omega)^\complex~\longrightarrow~{\rm
  G}\,/\,{\rm G}^\omega$ are trivial.
\label{buntrivlem}\end{theorem}
\noindent
The proof of Theorem~\ref{buntrivlem} is rather technical and can be
found in~\cite{AFQS1}.
\begin{proposition}
There is a natural algebra isomorphism
$$
\alg\bigl(W_{\mathcal{V}^\omega}\bigr)~\cong~\bigl\{f
\in C\bigl({\rm G}\,,\,{\rm End}\,\mathcal{V}^\omega\bigr)~\big|~
f(g\,h)=\mathcal{V}^\omega(h)^{-1}\,f(g)\,\mathcal{V}^\omega(h) \ ,
{}~ g\in{\rm G}\,,\,h\in{\rm G}^\omega\bigr\} \ .
$$
\label{Gmoduleprop}\end{proposition}
\begin{proof}
The vector space $C({\rm G})\otimes{\rm End}(\mathcal{V}^\omega)\cong
C({\rm G},{\rm End}\,\mathcal{V}^\omega)$ of matrix-valued functions
on $\rm G$ carries a natural $({\rm G}\times{\rm G}^\omega)$-action
$({\rm G}\times{\rm G}^\omega)\times C({\rm G},{\rm
  End}\,\mathcal{V}^\omega)\to C({\rm G},{\rm
  End}\,\mathcal{V}^\omega)$ given by
$$
\bigl((g,h)\,,\,f(g'\,)\bigr)~\longmapsto~
\mathcal{V}^\omega(h)\,f\bigl(g^{-1}\,g'\,h\bigr)\,\mathcal{V}^\omega
(h)^{-1} \ .
$$
This leaves an action of $\rm G$ on the space
$\alg(W_{\mathcal{V}^\omega})$ of ${\rm G}^\omega$-invariants. The
$\rm G$-module $\alg(W_{\mathcal{V}^\omega})$ can thereby be realized
explicitly in terms of ${\rm G}^\omega$-equivariant functions on $\rm
G$ as claimed.
\end{proof}

To construct gauge theory on the trivial rank~$1$ module over the
algebra $\alg(W_{\mathcal{V}^\omega})$, let $T^a$,
$a=1,\dots,\dim({\rm G})$ be a basis of generators of the Lie algebra
$\mfg$
obeying
\beq
\bigl[T^a\,,\,T^b\bigr]=\sum_{c=1}^{\dim({\rm G})}\,C^{abc}~T^c \ .
\label{Liealgrels}\eeq
Using the $\rm G$-action on the algebra $\alg(W_{\mathcal{V}^\omega})$
given by Proposition~\ref{Gmoduleprop} and the exponential mapping
$\exp:\mfg\to{\rm G}$, we define Lie derivatives
$L_a(f)$ of functions $f\in\alg(W_{\mathcal{V}^\omega})$ by
\beq
\bigl(L_a(f)\bigr)(g):=\frac{\dd}{\dd t}f\bigl(\exp(-t\,T^a)\,g\bigr)
\Big|_{t=0} \ , \quad a=1,\dots,\dim({\rm G})
\label{LiederivGdef}\eeq
which as vector fields on $\mfg$ obey the same Lie algebra relations
(\ref{Liealgrels}). As in Section~\ref{DbranesS3}, the natural
string-inspired gauge theory action functional
$S_\omega:\alg(W_{\mathcal{V}^\omega})\to\real$ reads~\cite{AFQS1}
\beq
S_\omega(A)=\int_{\rm G}\,\dd\mu_{\rm G}~\sum_{a,b=1}^{\dim({\rm G})}\,
\Tr\Bigl(\bigl(F_{ab}\bigr)^2+2\,\sum_{c=1}^{\dim({\rm G})}\,
C^{abc}~{\rm CS}_{abc}(A)\Bigr)
\label{SomegaAdef}\eeq
with all objects defined in a completely analogous way to those of
Section~\ref{DbranesS3}. Here $\dd\mu_{\rm G}$ is the invariant Haar
measure on the Lie group $\rm G$. The classical solutions of this gauge theory
again describe condensation processes on a configuration of D-branes
which drive the entire system into another D-brane
configuration~\cite{AFQS1}.

\subsection*{Acknowledgments}

The author thanks the organisors and participants of the workshop for
having provided a pleasant and stimulating atmosphere. In particular
he thanks J.~Cuntz, D.~Husemoller, M.~Khalkhali and M.~Sheikh-Jabbari
for enjoyable discussions. This work was supported in part by PPARC
Grant PPA/G/S/2002/00478 and by the EU-RTN Network Grant
MRTN-CT-2004-005104.

\end{document}